\newif\ifsubmode
\submodefalse

\newif\ifprintfig
\printfigtrue

\documentclass{aastex}

\shorttitle{$N$-band observations of He 2-10}
\shortauthors{Vacca et al.}

\received{}
\accepted{ }
\slugcomment{Accepted by AJ}

\newcommand{\mm}{$\rm \mu m$~}

\newcommand{\HII}{H\,{\small{\sc II}}}
\newcommand{\UDHII}{UD{H\,{\small{\sc II}}}}
\newcommand{\UCHII}{UC{H\,{\small{\sc II}}}}

\newcommand{\lta}{\lesssim}
\newcommand{\gta}{\gtrsim}

\begin{document}


\title{$N$-band Observations of He 2-10: Unveiling the
Dusty Engine of a Starburst Galaxy
\footnote{Based on observations obtained at the Gemini
Observatory, which is operated by the Association of Universities for
Research in Astronomy, Inc., under a cooperative agreement with the
NSF on behalf of the Gemini partnership: the National Science
Foundation (United States), the Particle Physics and Astronomy
Research Council (United Kingdom), the National Research Council
(Canada), CONICYT (Chile), the Australian Research Council
(Australia), CNPq (Brazil) and CONICET (Argentina)}}

\author{William D. Vacca}
\affil{Max-Planck-Institut fuer extraterrestrische Physik, Postfach 1312, 
D-85741 Garching, Germany}
\email{vacca@mpe.mpg.de}

\author{Kelsey E. Johnson }
\affil{JILA, University of Colorado and National Institute for 
Standards and Technology; and Department of Astrophysical and 
Planetary Sciences, Boulder, CO 80309-0440}
\email{kjohnson@colorado.edu}

\and

\author{Peter S. Conti}
\affil{JILA, University of Colorado and National Institute for 
Standards and Technology; and Department of Astrophysical and 
Planetary Sciences, Boulder, CO 80309-0440}
\email{pconti@jila.colorado.edu}


\ifsubmode\else
\clearpage\fi


\ifsubmode\else
\baselineskip=14pt
\fi


\begin{abstract} 
As part of an ongoing program to better understand the early stages of
massive star cluster evolution and the physical conditions for their
formation, we have obtained $J$, $K'$, and $N$ (10.8~$\mu$m) images 
of the nuclear region of the starburst galaxy He~2-10. The $N$-band images 
were obtained with the Gemini North telescope. In only ten minutes of 
on-source integration time with Gemini we were able to detect three of 
the five enshrouded clusters, or ``ultradense {\HII} regions'' ({\UDHII}s) 
recently discovered in radio maps. None of these sources appears in either 
the optical HST images or the near-infrared ($J$, $H$, and $K'$) images. These 
sources comprise about 60\% of the total $N$-band flux from He 2-10 and, 
we suspect, a similar fraction of the total far infrared flux measured 
by IRAS. 

The inferred spectra of the {\UDHII}s are strikingly similar to those
of Galactic ultracompact H {\sc ii} regions. We have modeled the radio+IR 
spectrum of these {\UDHII}s under the assumption that they are ``scaled-up" 
Galactic ultracompact H {\sc ii} regions.  From this model, the bolometric 
luminosity of the brightest cluster alone is estimated to be 
$\approx 2 \times 10^9$~L$_\odot$. 
The total mass of the dust and gas in this {\UDHII} is 
$M_{\rm shell} \approx 10^7~M_\odot$. 
We have also used the observed spectra to place constraints on the masses 
and ages of the stellar clusters enshrouded within the {\UDHII}s. For the brightest
{\UDHII}, we find
that the stellar mass must be $M_{\rm Cluster} \gta 2.5 \times 10^6 M_\odot$ 
and the age must be $\lta 4.8 \times 10^6$ yr, with the most probable age 
$\lta 3.6 \times 10^6$ yr. If we assume that the region 
is pressure confined and enforce the requirement that the star formation 
efficiency must be less than $\sim 90$\%, we find that the age of this stellar cluster 
must lie within a very narrow range, $4 \times 10^5 < \tau < 5 \times 10^6$ yr.
All of the clusters within the {\UDHII}s in He 2-10 are estimated to have ages less 
than about $5 \times 10^6$ yr and masses greater than about $5 \times 10^5  ~{\rm M}_\odot$.

We find that the logarithmic ratio of the radio to far-infrared flux 
densities, $q$, for the {\UDHII}s in He 2-10 is $\sim 4$; $q \approx 2.6$ for both 
He 2-10 as a whole and NGC 5253, another nearby starburst
known to host {\UDHII}s. These values of $q$ are significantly larger than the 
average $q=2.35$ found for normal galaxies, but comparable to the
values of $q$ found for ultraluminous infrared galaxies.
We suggest that large $q$ values for starburst galaxies may indicate that a significant 
fraction of the far-infrared flux may arise from thermal dust emission from {\UDHII}s.

Finally, the possibility that all of the far-infrared flux from He 2-10 and
other starburst galaxies may be produced by regions completely obscured at
wavelengths as long as $K'$ suggests that the well-known correlation between
ultraviolet continuum slope and infrared-to-ultraviolet flux ratio in
starbursts cannot be due entirely to reprocessing of ultraviolet
radiation by dust in a foreground screen geometry. In fact, the dust which reddens 
the ultraviolet continuum slope must be
largely decoupled from the dust which produces the large infrared fluxes in
some starbursts.
\end{abstract}




\section{INTRODUCTION}
Images of starburst galaxies obtained with the {\em Hubble Space Telescope 
(HST)} have resolved the sites of active star formation into numerous 
compact star clusters (see e.g., Schweizer 1999; Whitmore 2000 and references 
therein). These so-called ``super star clusters'' (SSCs) comprise
approximately 20\% of the total observed ultraviolet flux, and therefore 
represent a major mode of star formation in starburst galaxies (Meurer et al.\ 1995).
The radii of these SSCs are typically a few pc and masses, estimated from
both theoretical mass-to-light ratios and from direct measurements
of velocity dispersions, are in the range $10^4 - 10^6~M_\odot$.
The oldest of these SSCs have ages significantly greater than their 
dynamical timescales and hence must be bound. The masses, sizes, stellar
densities, and ages of SSCs suggest that the most massive examples may
be the progenitors of globular clusters. SSCs may also be intimately related
to nuclear stellar clusters recently found in the centers of exponential
bulges in spiral galaxies (Carollo, Stiavelli, \& Mack 1998; Carollo 1999; 
B\"oker, van der Marel, \& Vacca 1999; B\"oker et al.\ 2001).
Although the processes involved and the conditions required 
for massive star cluster formation are not completely understood,
it seems clear that extreme environments are necessary.  In the local 
universe vigorous star formation activity, for which SSCs are the signposts, 
often appears to be associated 
with cataclysmic events, such as galaxy mergers and interactions.

With ages $\tau > $ few Myr, the young blue SSCs discovered with HST 
(Whitmore 2000) have already emerged from their
birth clouds. Even earlier stages of cluster evolution must be hidden
from optical and ultraviolet (UV) observations by large amounts of molecular clouds
and dust from which the clusters are formed. Hence, studies of the early 
phases of cluster evolution require observations at longer wavelengths 
which are relatively immune to the effects of dust scattering and absorption 
by the natal material.

Recent radio observations of the starburst galaxies He 2-10 by Kobulnicky \& 
Johnson (1999, hereafter KJ99), and NGC 5253 by Turner and collaborators (e.g., 
Turner, Beck, \& Ho 2000; Gorjian, Turner, \& Beck 2001) suggest that examples 
of this young, embedded phase of cluster evolution may now have been identified. 
The radio maps of these starbursts reveal several compact thermal continuum 
sources that are optically thick at $\sim 6$ cm and longer wavelengths. 
While the radio spectra (either flat or rising with frequency) clearly indicate 
the emission from these sources is thermal in origin, the small sizes ($r \sim$ 
few pc), large inferred densities ($n_e \sim 10^3 - 10^5$ cm$^{-3}$) and enormous 
emission measures (${\rm EM} \sim 10^6 - 10^9$ cm$^{-6}$ pc) suggest however that 
these are not simply ``normal'' H {\sc ii} regions. Although H {\sc ii} regions 
with similar densities, emission measures, and so-called ``inverted spectra'' 
($S_\nu \propto \nu^{\alpha}$ with $\alpha > 0$)
exist in the Galaxy, and are known as ultracompact {\HII} regions (UC{\HII}s; 
for a review see Churchwell 1990, 1999), Galactic UC{\HII}s exist on vastly smaller 
mass scales: typical Galactic UC{\HII}s are excited by a single (or only a few) 
O-type star(s) (e.g., Churchwell, 
Wolfire, \& Wood 1990). The compact radio sources found in He 2-10 and NGC 5253, 
on the other hand, have inferred masses of $M \sim 10^5~M_\odot$, ionizing 
luminosities $N_{Lyc} \sim 10^{51 - 53}$, and contain the equivalent of 
$10^2 - 10^4$ O-type stars. The properties of these compact sources are clearly 
similar to those of the SSCs seen in the optical and UV images.
Because of the similarity of their radio spectra to those of UC{\HII}s, KJ99 
dubbed these objects ``ultradense {\HII} regions'', or UD{\HII}s. 

The existence of such dense thermal emission regions in starburst galaxies 
had been suggested prior to the recent results on He 2-10 and NGC 5253. 
Puxley et al.\ (1991) inferred the presence of compact, high density clusters
in NGC 2146 based on H53$\alpha$ observations. This result was confirmed by
Tarchi et al.\ (2000), who found six {\UDHII}s in 1.6 and 5 GHz maps of this
galaxy. Zhao et al.\ (1997), using H92$\alpha$ observations, found similarly 
dense regions in NGC 3628 and IC 694. Ulvestad \& Antonucci (1997) reported 
a number of such flat-spectrum sources in multi-wavelength radio maps of NGC 253. 
The unusual properties of the {\UDHII} in NGC 5253 were suggested by the 
work of Beck et al.\ (1996) and Turner, Ho, \& Beck (1998). In addition, the
VLA survey of so-called Wolf-Rayet galaxies by Beck, Turner, \& Kovo (2000)
indicates that {\UDHII}s are quite common in starbursts.

The significance of the recent results on He 2-10 and NGC 5253 lies not simply in
the detection of the {\UDHII} regions, but also in the morphologies of the 
radio maps relative to high spatial resolution optical images obtained with
{\it HST}. {\em None} of the radio ``knots'' detected in He 2-10 and NGC 5253 is 
associated with the SSCs readily seen in the {\it HST} ultraviolet and 
optical images. It is this detailed positional comparison, impossible before the 
launch of {\it HST}, that is so striking and that lends support to
the suggestion by KJ99 that these radio sources are in fact heavily dust-enshrouded
stellar clusters, the youngest phases of SSCs still buried deep
within their birth clouds. 

If these {\UDHII}s are indeed young SSCs hidden at ultraviolet and optical 
wavelengths as the result of absorption by surrounding dust, the bulk of 
their substantial bolometric luminosities should be radiated in the mid- to 
far-IR regimes as a result of the reprocessing of the cluster ultraviolet flux by 
the dust. With this in mind, we have carried out a set of near-infrared and thermal 
infrared imaging observations in an attempt to detect the {\UDHII}s directly from 
their dust emission. In this paper we present the detection of thermal-infrared 
emission in the $N$-band from the {\UDHII}s discovered by KJ99 in He~2-10.
The observations are described in \S 2. A discussion of the relative morphologies, an
analysis of the spectral energy distributions (SEDs) of the radio knots, and
constraints on the properties of the embedded stellar clusters are presented in \S 3. 
We compare the properties of the {\UDHII}s seen in He 2-10 with those of Galactic
{\UCHII}s and discuss the implications our results have on the infrared-radio
correlation and the $F_{\rm FIR}/F_{1600}$ correlation for starburst galaxies in
\S 4. Our conclusions are summarized in \S 5.

\section{OBSERVATIONS}
\subsection{$N$-band images}
$N$-band ($\lambda_{avg} \sim 10.8~\mu$m) observations of He 2-10 were made using 
the mid-infrared imager OSCIR on the Gemini North telescope during the scheduled 
OSCIR mini-queue in February 2001. OSCIR contains a $128 \times 128$ Si:As BIB 
detector sensitive between 8 and 25 $\mu$m. At Gemini North, the field of view of 
OSCIR is $11.''39 \times 11.''39$ and the pixel scale is $0.''089$ pix$^{-1}$.
Conditions were photometric during the observations and the seeing, as measured by 
the FWHM of standard stars, was in the range of $0.''8~-~0.''9$. However, due to high winds, 
unfortunately only a total of ten minutes of `on source' integration time were 
possible.  Two sets of 5 minute observations (with small offsets in pointing) were 
made of He~2-10 at an airmass of $\sim 1.9$.
Throughout the observations, standard chopping and nodding
techniques were employed in order to subtract the thermal sky background.  

The images were reduced using the {\small OREDUCE} task in the {\small
GEMINI/OSCIR} package available in the Image Reduction and Analysis
Facility (IRAF). The two sets of images of He 2-10 were reduced separately
and then shifted (to account for the pointing offsets) and combined. The
final $N$-band image, corresponding to $9.''43 \times 9.''43$, 
is shown in Figure ~\ref{Nimage}, which reveals several
marginally-resolved, closely-spaced emission regions. Photometry of these
regions was performed by constructing a model of the source, consisting of 
tilted, elliptical Gaussians atop a constant background, and fitting it to 
the image using a least-squares routine. In addition to providing estimates 
of the total fluxes and sizes of the extended emission regions, and a 
goodness-of-fit value, this method also ensures consistency between the $N$-band 
fluxes and the radio fluxes, which were derived in a similar fashion. 

The data were calibrated using observations of the standard stars
$\alpha$~CrB, $\beta$~Leo, and $\gamma$~Pix. We adopted the $N$-band magnitudes
given by Tokunaga (1984) for $\alpha$~CrB and $\beta$~Leo. The $N$-band
magnitude of $\gamma$~Pix was calculated using the absolutely 
calibrated spectrum given by Cohen et al.\ (1999) and the OSCIR filter
transmission curve. We adopted the median $N$-band extinction coefficient for Mauna Kea
given by Krisciunas et al. (1987) and solved for the photometric zero point.
The zero point and the extinction coefficient were then applied to the 
instrumental fluxes computed for the various emission regions in He 2-10. 
Flux densites in mJy were computed using the conversion given on the 
OSCIR web page (0 mag $= 37.77$ Jy). The absolute uncertainties in the source 
fluxes are $\sim~15-20$~\% and are dominated by the uncertainties in the 
photometric calibration derived from the standard stars.  

\subsection{Near-infrared images}
As part of a larger project to image starburst galaxies in the near-infrared,
He 2-10 was observed in the $J$, $H$, and $K'$ bands using the University of 
Hawaii 2.2m telescope with the Quick Infrared Camera (QUIRC, Hodapp et 
al.\ 1996 ) and the UH tip/tilt system (Jim et al.\ 2000) on 02 January 1996.
QUIRC contains a $1024\times1024$ HAWAII array produced by Rockwell and is sensitive
between 1 and 2.5$\mu$m . At the $f/31$ focus of the 2.2m telescope, the
pixel scale is $0.''06$ pix$^{-1}$ and the field of view is approximately 
$1'\times 1'$. During the observations, conditions were photometric and the 
seeing, as measured from standard stars, was about $0.''6$ at $K'$.

Standard procedures were used for observing an extended source in 
the near-infrared (dithering and nodding to sky). A set of 7 on-source
images, each 180 s in $J$ and $H$ and 90 s in $K'$, were acquired. A set of
equal length sky exposures were interleaved between the source frames. The images
were reduced in IRAF using a suite of specially designed routines.
After subtracting the sky and flat-fielding, the images were aligned
and combined using a median routine. The $K'$ image of the center of He 2-10,
with the central position, size, and orientation corresponding to the $N$-band image,  
is presented in Figure ~\ref{Kimage}; the full $J$ band image of He 2-10 
was presented by Vacca (1997).  Flux calibration was determined 
from near-infrared photometric standard stars (Casali \& Hawarden 1992)
observed during the night.
A detailed discussion and analysis of the near-infrared images will be
presented elsewhere (Vacca, in preparation).
  
\section{RESULTS}
\subsection{$N$-band Morphology}

Figures \ref{Nimage} and \ref{Kimage} show the $N$ and $K'$ images 
of He 2-10, respectively; the orientations and fields of view of these images 
are the same. For an assumed distance to He 2-10 of 9 Mpc (Vacca \& Conti 1992), 
$1''$ corresponds to about 44 pc; the full image 
shown in Figure ~\ref{Nimage} covers approximately 410 pc $\times$ 410 pc. Despite 
the lower resolution of the ground-based observations, the 
$K'$ image appears very similar to the HST UV and optical images presented
by Conti \& Vacca (1994) and Johnson et al.\ (2000), in that it reveals a
bright arc of emission, and a luminous point source at the top of the arc 
corresponding to the brightest UV/optical starburst knot, 
against a diffuse background. 
A detailed comparison between the near-IR images and the HST images 
will be presented elsewhere (Vacca, in prep.). 
The $N$-band image, however, which reveals three bright sources atop a much weaker 
background, is striking for its {\em dissimilarity} to either the 
optical or near-infrared images of this galaxy. 
To demonstrate this, in Figure \ref{Noptical} we show the contours of the optical 
V-band image from HST obtained by Johnson et al.\ (2000) on top of the $N$-band image. 
Even with the relative astrometric uncertainty in the HST image of $\sim 1''$, 
it is clear that {\em none of the SSCs seen in the HST UV and optical images, or 
the ground-based near-infrared images, is detected in the Gemini $N$-band image.} 
The starburst clusters seen in the optical image are not associated with the bright 
mid-IR emission regions. In fact, it appears that the bright $N$-band regions serve
to delimit the extent of the optical emission.
The arc of UV/optical starburst knots is entirely located  
between the two brightest $N$-band sources directly to the east and west;
the brightest UV/optical SSC is located at the position of the weakest $N$-band
emission between these two bright sources. 
This region of weak 10.8 $\mu$m emission also shows very low H$\alpha$ equivalent 
width, which Johnson et al.\ (2000) suggest is due to the strong winds from the starburst
knots having blown away the ambient gas. The $H\alpha$ equivalent width map
shows no other obvious correlation (or anti-correlation) with the $N$-band image; 
weak $H\alpha$ emission is seen throughout the region where the brightest $N$-band 
source is located, for example. 

In contrast with the UV, optical, and near-infrared images, the radio maps of KJ99 
are extremely well-correlated with the $N$-band image, as shown in Figure~\ref{Nradio}, 
which presents the 2 cm flux contours on top of the $N$-band image. The four
sources detected in the $N$-band are clearly associated with four
or five of the radio-bright {\UDHII}s found by KJ99.
{\it The morphology of the mid-IR emission is nearly identical to that of 
the thermal radio emission and is not correlated with the UV, optical, or near-IR 
morphology}. This result immediately suggests that most of the 10.8 \mm flux
is not due to ambient dust that has been heated (primarily) by the starburst knots 
seen in the UV/optical. As we will demonstrate below (\S 4.1), 
heating of dust in the bright 10.8 \mm regions and their surrounding vicinity 
by these optical SSCs is negligible. Rather, 
the 10.8 \mm emission arises primarily from hot dust heated by the stars embedded
within the {\UDHII}s themselves.

In KJ99 the {\UDHII}s are numbered 1-5 from west to east 
in He~2-10.  Using this nomenclature, we have detected the {\UDHII}s 1, 3, 4, and 5 
at 10.8~$\mu$m; {\UDHII} \#2 is likely to be blended with \#1 in the $N$-band image. 
All the knots have observed FWHM values larger than the seeing, and are therefore 
marginally resolved. Knot 4, in particular, is clearly extended in one direction, 
in agreement with the morphology seen in the radio maps. This may be due to the 
presence of several sources which are not separated at our resolution. 
The fluxes and $N$-band magnitudes of the four $N$-band knots are listed in 
Table~\ref{flux.tab} 
along with the effective radii derived from the elliptical Gaussian fits used 
to determine the photometry. The effective radii were calculated from
\begin{equation}
R_{\rm eff} = a \cdot \sqrt{\epsilon}
\end{equation}
where $a$ is the semi-major axis and $\epsilon$ is the axial ratio determined 
from the fits. We corrected these values for broadening due to Gaussian seeing. 

The total $N$-band flux found from our image is $720 \pm 95$ mJy.
This value is in reasonable agreement with that given by Telesco, Dressel, 
\& Wolstencroft (1993), who measured the integrated $N$-band flux of He~2-10 
to be $600 \pm 40$~mJy. Both of these values, however, are substantially smaller 
than those found by Sauvage, Thuan, \& Lagage (1997; 1.06 Jy) and Cohen \& Barlow 
(1974; 1.6 Jy). 
\footnote{After this paper was submitted the recent $11.7 \mu$m observations of 
He 2-10 by Beck, Turner, \& Gorjian (2001) were brought to our attention. The 
morphology of He 2-10 at $11.7 \mu$m appears to be very similar (as expected) to 
that seen in our $10.8 \mu$m image. Beck et al.\ (2001) measure a total flux at 
$11.7 \mu$m of 880 mJy, which represents at least 80\% of the total 12 $\mu$m 
flux measured by IRAS. This value is in reasonable agreement with our estimate 
of the total flux at $10.8 \mu$m, given that the infrared spectrum of He 2-10 as 
measured by IRAS is steeply rising between 12 and 25 $\mu$m; our
value is within the 20\% uncertainty of theirs. However, 
their estimates of the fluxes for the individual knots seen in the images are about a 
factor of 2 larger than ours. This is presumably due to the fact that Beck et al.\ 
(2001) did not account for the background beneath the knots in their photometric 
measurements. Nevertheless, the fractional contributions of the knots to the total
$11.7 \mu$m flux are identical to those found here. The physical sizes of the knots
as estimated by Beck et al.\ (2001) are somewhat smaller than, although comparable to, 
those we have derived.} 

Sauvage et al.\ (1997) dismissed the measurement of Cohen \& Barlow 
(1974) as being in error since it is substantially larger than the 12 \mm IRAS
flux. The difference between our flux value and that of Sauvage et al.\ (1997)
is more puzzling, however,
as He 2-10 is a fairly compact source and
all of the flux contours presented by Sauvage et al.\ (1997) are contained
within the field of view of our image. In fact, their deconvolved image appears to
be nearly identical to ours, and all of their measured flux is contained within
the same regions we detect (see their Table 1).
Therefore, it is unlikely that Sauvage et al.\ (1997) could have detected significantly
more flux from He 2-10 (in, for example, a faint diffuse extended component) than 
we see. 

The discrepancy is probably due to three factors: (1) large differences in the 
$N$-band filter transmission profiles and central wavelengths; (2) an attempt by 
Sauvage et al.\ (1997) to incorporate a color term in their photometric calibration 
(we have not included a correction for the color differences between our standards 
and He 2-10); and (3) non-photometric conditions during the observations of Sauvage 
et al.\ (1997). With regard to point (2), we should point out that, while there 
probably is a potentially large color term in the photometric transformations between 
the standards and He 2-10, the color term correction adopted by Sauvage et al.\ (1997) 
is almost certainly incorrect. Sauvage et al.\ (1997) extrapolated their observed 
10.1~$\mu$m to the $N$-band assuming a spectral energy distribution
of $f_\nu \propto \nu$; however, as we show below, the spectral energy 
distribution of He 2-10 near $10\mu$m is actually decreasing rapidly
with frequency (see also Zinnecker 1987). Additionally, if we fit a blackbody spectrum 
to the IRAS 12~$\mu$m and 25~$\mu$m flux points, 
and extrapolate to $10.8~\mu$m, we predict a total $N$-band flux of
640 mJy, in good agreement with both our observed value and that of Telesco et al.\ (1993).
It should also be noted that Sauvage et al.\ 
(1997) assumed the thermal-IR emission is coincident with the optically visible 
starburst.  As can be seen from Figures~\ref{Nradio} and \ref{Noptical}, the dust 
emission coincides with the thermal radio morphology
and is {\it not} associated with the optically visible SSCs.

The combined $N$-band flux of the four detected {\UDHII}s indicates that
$\sim 60$\% of the $10.8~\mu$m flux from the entire galaxy can be attributed 
to just these four objects. Almost all of the remaining $N$-band 
flux is due to the background on which the {\UDHII}s sit, and which has a morphology 
very similar to that of the {\UDHII}s, but on a slightly more extended scale. 
Very little flux arises from any diffuse component beyond the immediate location 
of the {\UDHII}s. The surface brightness of the background beneath the {\UDHII}s 
is approximately 10 mJy/sq arcsec. Based on the relative morphologies of the 
HST images and the background detected in the $N$-band image, it would also appear that 
very little of this background flux is due to emission from dust directly surrounding 
and heated by the SSCs seen in the UV and optical images. 
It is possible that some of the background flux 
arises from dust which is heated by radiation escaping from the {\UDHII}s. If
this is indeed the case, then nearly all of the observed $N$-band flux 
and, by extension nearly all of the observed flux in the 12 \mm IRAS band, from He 2-10
would be attributable to {\UDHII}s that are invisible in UV, optical, and 
near-infrared images. Given the vigorous star formation throughout
He 2-10, as traced by the HST UV and optical images (Conti \& Vacca 1994;
Johnson et al.\ 2000), this result is in itself rather remarkable.

The relative locations along the line of sight of the $N$-band knots and the 
optically bright starburst knots are not known; the optical SSCs 
could be located in front, behind, or at the same distance as the $N$-band sources.
If the optical SSCs are located behind the $N$-band knots, this would
suggest that the observed optical morphology is completely determined and 
delineated by the foreground regions of low dust optical depth (and hence
weak or absent $N$-band emission).\footnote{This is the ``hole story'' scenario
of Meurer et al.\ (1995).} On the other hand, if the $N$-band knots and the
optical SSCs reside at the same relative distance, it is tempting to
conclude that the $N$-band morphology may, to some extent, be a result of the
sweeping action of the wind arising from the optically visible starburst.
The $N$-band emission would then identify regions where dust has been
swept up, the density has increased, and the resulting structures have
collapsed and begun to form stars; the areas of faint or absent $N$-band 
emission would indicate regions 
where the dust has been effectively cleared out by the starburst wind.   

\subsection{Observed Spectral Energy Distributions and Physical Models for the {\UDHII}s}

We have constructed an SED for each of the four $N$-band knots by combining the measured
$N$-band fluxes, the radio fluxes at 2 and 6 cm from KJ99, additional fluxes at 1.3 cm
for three knots (kindly provided by H.\ Kobulnicky, private communitcation),
and estimated fluxes in the
IRAS bands. The last were estimated as follows. As discussed above, there is a 
good correspondence between the total $N$-band flux observed from He 2-10 and that 
predicted at 10.8 $~\mu$m by fitting the short wavelength IRAS fluxes with a simple 
blackbody. Hence, it appears that the knots seen 
in the $N$-band image can account for most of the $12\mu$m IRAS flux from He 2-10. 
\footnote{This has been confirmed by the $11.7 ~\mu$m observations of He 2-10 by
Beck et al.\ (2001). See footnote 3 and the previous section.}
If we now
assume that the $N$-band and radio images are representative of the spatial distribution
of the emission regions detected in the IRAS bands, we can then infer the 12, 25, 60, and 100
$\mu$m fluxes for each of the knots by multiplying the total IRAS fluxes for He 2-10 
at these wavelengths by the fractional
contributions of each knot to the total $N$-band flux. 
The resulting SEDs are 
shown in Figure~\ref{SEDs} and bear a remarkable resemblance to the SEDs for 
Ultracompact H {\sc ii} regions (e.g., Wood \& Churchwell 1989).

Using the analysis of {\UCHII} regions as a guide (see e.g., Churchwell 
et al.\ 1990), we have constructed simplified physical models of
{\UDHII} regions and attempted to match the observed SEDs of the knots with 
the model spectra. The models consist of a ``cocoon'' comprising two emission 
zones: an inner spherical region ($0 < r < R_{\rm in}$),
which produces the thermal bremsstrahlung radio spectrum, surrounded by a spherical
dust shell with $R_{\rm in} < r < R_{\rm out}$, which is responsible for
most of the far-infrared emision. The entire cocoon surrounds a stellar cluster
which provides the photons necessary to ionize the inner region and heat the
dust in the outer shell. For the free-free emission region
we adopted the mean electron temperature found by Kobulnicky et al.\
(1999) of $T_e = 6000$ K, and assumed a constant electron density $n_e$. 

For simplicity, we assumed the dust consists of spherical silicate grains 
with an opacity proportional to $\nu^{1.5}$ and a radius of 0.1 \mm . The
absorption efficiency $Q_{abs}$ was normalized to a value of 0.01 at 40
$\mu$, as given by Draine \& Lee (1984).
The dust density was assumed to be constant with shell radius. 
The latter assumption has been shown to provide a good
fit to the spectra of Galactic {\UCHII} regions (Churchwell et al.\ 1990)
The dust shell was assumed to be optically thin at all wavelengths.
The far infrared data ($N$-band and IRAS points) clearly indicate that a single
temperature blackbody is not a good representation of the SED.
Therefore, the dust temperature within
the shell was allowed to vary with radius between values specified
at the inner and outer edges of the shell, $T_{\rm in}(R_{\rm in})$
and $T_{\rm out}(R_{\rm out})$, respectively. We adopted the specific functional
form $\log T \sim (r/R_{\rm in})^{-\delta}$, where $\delta$ was determined from the fit
to the data. This form was adopted primarily because it is fairly general and can reproduce
the qualitative characteristics of the temperature profile found by Churchwell et al.\ (1990)
in their analysis of the {\UCHII} G5.89-0.39. We made no attempt to determine 
the temperature self-consistently, however, and therefore these
models are not completely realistic; nevertheless, they are useful for
obtaining rough estimates of the physical parameters of the knots.
 
The model was fit to the SEDs of each of the knots using a least-squares 
procedure. The relevant parameters resulting from this procedure are 
given in Table ~\ref{pars}. 
The radii $R_{\rm in}$ and the electron densities $n_e$
we find are similar to those derived by KJ99; the large electron densities
are required to produce the turnover observed in the SEDs at the lowest frequencies.
Because of the manner in which the fluxes in the IRAS bands were estimated,
the derived dust shell parameters are all very similar for the four knots.
Figure ~\ref{SEDmodel} shows the resulting best fit to the spectrum of knot 4. 
The temperature profile as a function of shell radius for knot 4 is shown in
Figure ~\ref{templot}. As this figure shows, the dust
temperature varies between $T_{\rm in}(R_{\rm in}) > 180~$K at the
inner edge and $T_{\rm out}(R_{\rm out}) = 29~$K at the outer edge,
falling off quickly with radius; $\delta$ was found to be $\sim 0.11$,
and hence the temperature drops from its maximum value to below $100$ K within
a radius of 10 pc, or less than 0.1\% of the total shell volume. 
Throughout most of the shell the dust temperature is between $50$ K and $29$ K.
The outer temperature $T_{\rm out}(R_{\rm out}) = 29~$K is in good
agreement with the value expected for the ambient interstellar medium.
Churchwell et al.\ (1990) found similar results in their more complete 
models of {\UCHII} regions. Despite the large outer radius, it is 
not in contradiction with the observed sizes, as the half-light radius 
estimated from the temperature distribution is only about a few pc.
Rather, the fact that the predicted half-light radii are so much smaller than 
the radii measured from the $N$-band image suggests that the completely
optically thin assumption we adopted may not be valid for the shorter
wavelengths (10.8, 12, and possibly 25 $\mu$m). 

The total far infrared luminosity between 10 \mm and 1 cm for knot 4 
is $1.8 \times 10^9 L_{\odot}$, with an uncertainty of about 20\%. (The
uncertainties on the infrared luminosities of the other knots are about 25\%.) 
This represents about $30$\% of the total far infrared flux from He 2-10.
The total mass found for the dust shell, 
$M_{\rm shell}$, is about $1.2 \times 10^7 M_{\odot}$, where we have 
assumed the standard dust-to-gas ratio of $0.007$ (Whittet 1992). The dust mass 
is highly uncertain, however, as it depends critically on the exact values of
the dust properties and the absorption coefficient. 
A significantly larger mass for the shell would result from using the 
substantially lower dust-to-gas ratio of $2-6 \times 10^{-4}$ suggested by
Baas, Israel, \& Koornneef (1994). 
Additionally, the total shell mass could well be larger than our estimate if the dust 
shell is actually optically thick at the shorter wavelengths. One indication
that this could indeed be the case (in addition to the half-light radii mentioned
above) comes from the estimate of the optical and near-infrared extinction.
Despite the fact that the {\UDHII}s are not detected in either our optical or 
near-infrared images, the estimated dust masses derived from our model fits correspond 
to only moderate optical extinction values of $A_{5000}$ between about 0.7 
(for knot 3) and 12 (for knot 4); the corresponding extinction values at 2.2 $\mu$m 
are estimated to be only between 0.05 and 0.9.
(For these estimates we have adopted the extinction laws appropriate for the 
silicate-graphite mix of Mathis, Rumpl, \& Nordsieck 1977, as calculated
by Draine \& Lee 1984). Clearly, there must be more dust than indicated by
our model fits or the knots would be visible in our $K'$-band image. 

We point out that we do not know the extent to which emission from fine structure 
nebular lines (e.g., [Ar {\sc iii}] 8.9~$\mu$m, [S {\sc iv}] 10.5~$\mu$m, and 
[Ne {\sc ii}] 12.8~$\mu$m) and polycyclic aromatic hydrocarbons (PAHs, at 8.6, 
11.3, and 12.6~$\mu$m) might be contaminating the observed $N$-band fluxes 
of the individual knots. (The filter transmission is $> 80\%$ between 8 and 13 
$\mu$m.) Starburst
galaxies are known to exhibit a wide range of strengths for these features (Madden
2000). We can estimate the possible nebular line contamination to the 
{\em total} $N$-band flux of He 2-10 using the [Ar {\sc iii}], [S {\sc iv}], 
and [Ne {\sc ii}] emission line strengths reported by Beck, Kelly, \& Lacy (1997).
The total flux from these three emission lines is $\sim 3 \times 10^{-12}$ erg
cm$^{-2}$ s$^{-1}$. The size of the region to which this total line flux corresponds 
is unclear from the Beck et al.\ (1997) paper, but probably includes all four
$N$-band knots. In this case, the emission line flux represents only $\sim 5-6$ \%
of the total measured $N$-band flux from the four knots, far less than our 
photometric uncertainty.
Therefore, we have ignored the possible nebular line contamination to the 10.8
$\mu$m fluxes. The possible PAH contribution is far more uncertain.
A mid-IR spectrum of the starburst galaxy NGC~5253 presented by Crowther 
et al.\ (1999), however, shows no sign of PAH emission. We have therefore 
assumed that these features also do not contribute substantially to the $N$-band flux
from He 2-10.

\subsection{Mass and Age of the Stellar Cluster in Knot 4}

Despite the limited knowledge of the SEDs of the $N$-band/radio knots, 
a number of constraints can be placed on the ages and 
masses of the stellar clusters embedded in these {\UDHII}s. We will 
demonstrate these constraints using knot 4 as an example.

Because the {\UDHII}s are not seen in the UV, optical, and near-infrared images, 
we can approximate the
bolometric flux for the knots as $L_{\rm bol} \sim L_{\rm IR}$. This
estimate is strictly a lower limit, as the knots could well be
emitting some (presumably small) fraction of $L_{bol}$ between 2.5 and
$10\mu$m. If we now make the usual assumption that the infrared flux
represents the reprocessed emission from a stellar cluster, we can 
compare our estimates of $L_{bol}$ with those predicted by 
population synthesis models to constrain the mass and age of the
cluster. We have used the updated version of the Bruzual \& Charlot
(1993) models (hereafter referred to as BC00) models with the
following parameters: (1) an instantaneous burst of star formation;
(2) solar metallicity; and (3) a Salpeter Initial Mass Function
between 0.1 and 100 M$_{\odot}$. For knot 4, the observed value of
$L_{\rm bol}$ can be reproduced by the model clusters with the set of 
ages and masses indicated by the solid line shown in the
age-vs.-mass plot in Figure~\ref{mtplot}. Because the estimated
value of $L_{\rm bol}$ is a lower limit to the true value, the region
below this line is excluded.

The relatively flat radio spectra ($\alpha^6_2 \sim 0.1-0.5$) of the knots 
between 5 GHz (6 cm) and 23 GHz (1.3 cm), and the observed turnover in the
spectra at low frequencies, are indicative of thermal free-free emission, as 
discussed by KJ99, and as we have assumed in our models (\S 3.2). 
Most galaxies, however, exhibit radio spectra that decrease with
increasing frequency, indicative of synchrotron emission from SNe and
cosmic rays. Indeed, He 2-10 as a whole also has a non-thermal radio
spectral index. The thermal nature of the radio spectrum of the knots can be used 
to place limits on the SN frequency, and consequently on the ages, of the knots. Let us
assume for the moment that {\it all} of the flux emitted at 6 cm is due to
synchrotron radiation; this is clearly an upper limit to the non-thermal flux. 
The relation between the non-thermal luminosity and the SN rate is given 
by Condon \& Yin (1990) and Condon (1992):
\begin{equation}
\frac{L_{\rm non-therm}}{10^{22} ~{\rm W~Hz^{-1}}} \sim 13 \Bigl(\frac{\nu}{\rm GHz}\Bigr)^{\alpha} 
\Bigl(\frac{\nu_{\rm SN}}{\rm yr^{-1}}\Bigr)~~.
\end{equation}
This equation can be used to derive an upper limit to the SN rate in knot 4,
once the 6 cm flux has been scaled to account for the possible effects of synchrotron 
self-absorption. Adopting a value of $\alpha \sim -0.8$ typical for Galactic synchrotron
sources (Condon 1992), we find $\nu_{\rm SN} \leq 2.1 \times 10^{-3}$ yr$^{-1}$.
Again employing the models of BC00, we use this value 
to place the additional constraints on the mass and age of the stellar cluster in
knot 4 shown in Figure~\ref{mtplot}. Regions above the designated 
line in this figure are excluded by the lack of any signatures of SNe
in the SED of knot 4.


The radio flux can also be used to estimate the total ionizing
flux from the stellar cluster. Assuming that all the radio flux is 
due to thermal bremsstrahlung, and correcting for self-absorption,
we find the total thermal luminosity for knot 4 is $L_{\rm therm} 
\sim 1.8-2.3 \times 10^{19} ~{\rm W~ Hz^{-1}}$,
depending on the wavelength used for the estimate.
This value corresponds to a limit on the total ionizing flux from the
cluster of $\log Q_0 ~({\rm s^{-1}}) > 52.3 ~,$ where we have used the
relation between the radio luminosity and the predicted number of ionizing photons 
emitted per second by a cluster given by Condon (1992)
\begin{equation}
\log Q_0 \gta 52.8 - 0.45 \log\Bigl(\frac{T_e}{10^4 {\rm K}}\Bigr) + 0.1 \log\Bigl(\frac{\nu}{{\rm GHz}}\Bigr) + \log\Bigl(\frac{L_{\rm therm}}{10^{20} ~{\rm W Hz}^{-1}}\Bigr)~.
\label{eq:q0}
\end{equation}
This value of $Q_0$ is a lower limit, as Equation \ref{eq:q0} does not account 
for the effects of absorption of ionizing photons by dust.
Of course, $Q_0$ could well be smaller than this estimate if a significant fraction of the 
radio flux is due to synchrotron emission rather than thermal bremsstrahlung. However, our
estimate agrees well with that determined by Mohan, Anatharamaiah, \& Goss (2001) from 
observations of the H92$\alpha$ emission line ($Q_0 > 2.5 \times 10^{52} ~{\rm s}^{-1}$). 
The limit on $Q_0$ corresponds in turn to limits on the mass and age of the
stellar cluster. From the BC00 models, we find that the estimated value of $Q_0$ can
be produced by stellar clusters with the ages and masses indicated by the designated
line in Figure~\ref{mtplot}. Regions below this line 
are excluded because too few ionizing photons would be produced to
account for the observed radio luminosity.

We can also apply lifetime arguments to the knots in He 2-10 to derive
additional constraints on the stellar cluster masses and ages, as well as the 
ambient density in the vicinity of the {\UDHII}s. The radius of an H {\sc ii} 
region should expand according to the relation given by Spitzer (1978),
\begin{equation}
\label{size}
r(t) = r_i [ 1 + (7c_st/4r_i)]^{4/7}~~~,
\end{equation}
where $c_s$ is the local sound speed, and $r_i$ is the initial size 
of the Stromgren sphere immediately after the ionizing stars ``turn on'',
given by
\begin{equation}
\label{rstrom}
r_i = \Bigl[\frac{3Q_0(1-f_d)}{4\pi n_0^2 \alpha_B}\Bigl]^{1/3}~~~,
\end{equation}
where $f_d$ is the fraction of ionizing photons absorbed by dust; 
we have assumed $f_d = 0$.

We can invert
this equation to determine the age and mass of the stellar cluster for
various values of the initial (assumed to be the ambient) density $n_0$. 
The results indicate that  
masses and ages consistent with the other constraints shown in Figure ~\ref{mtplot}
can found only if the ambient density is on the order of $n_0 \sim 10^5 - 10^7$
cm$^{-3}$.
We can estimate $n_0$ independently from equilibrium arguments. 
The H {\sc ii} region stops expanding 
when it comes into pressure equilibrium with the ambient environment, such that 
\begin{equation}
\label{dens}
P = \gamma n_e k T_e = n_0 kT_0 = P_0~~~,
\end{equation}
where the left hand side of the equation corresponds to the conditions within
the ionized region and the right hand side corresponds to the ambient environment.
The value of the numerical constant $\gamma$ depends on the dominant state of
the gas in the ambient medium. If the gas consists mainly of atomic hydrogen,
$\gamma = 2$; if  molecular hydrogen is the main component of the ambient medium,
$\gamma = 4$ (De Pree, Goss, \& Gaume 1998).
The final radius of the ionized region is given by 
\begin{equation}
\label{revol}
r_f = r_i \Bigl(\frac{Q_0(t=t_f)}{Q_0(t=0)}\Bigr)^{1/3} \Bigl(\frac{\gamma T_e}{T_0}\Bigr)^{2/3}~~~,
\end{equation}
where $t_f$ is the time at which equilibrium is reached and $Q_0(t=t_f)$ is the
ionizing flux produced by the cluster at this time.
Assuming that $n_e = 4290$ cm$^{-3}$ and $T_0 = 29~$K, as found from our model
of the SED of knot 4, and adopting $T_e = 6000~$K and $\gamma = 2$, we use 
Equation \ref{dens} to solve for $n_0$ and find $n_0 = 1.8 \times 10^6$
cm$^{-3}$. Given the location of the radio knots, very close to starburst regions
which are visible in the ultraviolet (and hence have blown away their birth clouds),
such densities do not seem unrealistic. They are also consistent with estimates of
ambient densities around Galactic {\UCHII} regions (e.g., De Pree, Rodr\'iguez, 
\& Goss 1995). Both the original ambient density and the wind arising from the optically 
visible starburst may serve to pressure confine the {\UDHII}s (and possibly extend 
their lifetimes). As suggested in \S 3.1, such a scenario would also be consistent with
the H$\alpha$ equivalent width map of He 2-10 presented by Johnson et al.\ (2000), as
well as with the relative morphologies of the $N$-band and UV/optical images.

Using this estimate of $n_0$, the current radius of the ionized region $r=R_{\rm in}=3~$pc as 
determined from the fit to the SED of knot 4, and the predictions of the BC00 models 
for the value of $Q_0$ as a function of the mass of a young stellar cluster, 
we derive the relation between the age and mass of the stellar cluster within knot 4 
plotted in Figure ~\ref{mtplot}. This relation is strictly a lower limit,
as both absorption of ionizing photons by dust ($f_d > 0$) and the use of
$\gamma = 4$ yield larger cluster masses for any given age. Hence, cluster age and mass
combinations below this line in Figure ~\ref{mtplot} are excluded.
Note, however, that the
estimate for $n_0$ explicitly assumes that the H {\sc ii} is in pressure equilibrium with
its surrounds. For the young ages that are possible for these {\UDHII}s this 
may not be valid. 


As an aside, we note that one of the outstanding problems in the study of UCHII 
regions has been their relatively long lifetimes. Early estimates yielded an 
expected lifetime of a typical {\UCHII} of only $10^4$ yr, far too short to 
account for the number of {\UCHII}s seen in the Galaxy (Wood \& Churchwell 1989). 
De Pree, Rodr\'iguez, \& Goss (1995; see also Garc\'ia-Segura \& Franco 1996), 
however, have suggested that the solution to this 
puzzle lies in the value of the ambient gas density in which an {\UCHII} is formed.
If the ambient density is on the order of $n_0 \sim 10^7$ cm$^{-3}$, rather than
$n_0 \sim 10^5$ cm$^{-3}$, the
estimated lifetime of an {\UCHII} can be increased by a factor of 10 or more
beyond the original value estimated by Wood \& Churchwell (1989).  
Given the observational similarities between {\UDHII}s and {\UCHII}s, we might 
expect these objects to be formed in analogous physical environments.  

As Figure~\ref{mtplot} demonstrates, the age of the stellar cluster in knot 4 
is constrained to be less than about $4.8 \times 10^6$ yr, with the most probable
value less than $3.6 \times 10^6$ yr, and the mass must be
larger than about $2.5 \times 10^6 ~{\rm M}_\odot$. Carrying out similar
calculations for the other knots seen in the $N$-band image reveals that they
all must have ages less than about $5 \times 10^6$ yr and masses greater
than about $5 \times 10^5  ~{\rm M}_\odot$.
Using the lower limit of the mass of the stellar cluster and the 
estimated mass of the surrounding dust shell for knot 4, we find a star formation 
efficiency of about 17\% for this knot. The uncertainties on this value are, of course, 
very large.

If we now assume that the total stellar mass in knot 4 must be less than about
$10^9~M_\odot$ [where we have increased the estimated cocoon mass by a factor of 10,
to allow for the possibilities that the SED model may have underestimated the 
total dust mass and that the dust-to-gas ratio may be smaller in irregular blue 
starburst galaxies (Hunter et al.\ 1986), and have required the star formation 
efficiency to be $< 90$\%], we find that the age of the cluster in knot 4 must 
lie within a narrow range between $4 \times 10^5 \lta \tau \lta 5 \times 10^6$ yr. 
Decreasing the mass of the stellar cluster raises the lower age limit and narrows 
the range of possible ages; decreasing the age increases the lower mass limit. 
The total dynamical mass of He 2-10 was
estimated by Kobulnicky et al.\ (1995) to be of the order of $5 \times 10^9~M_\odot$.
Therefore, it seems somewhat unlikely that $10^9~M_\odot$ of material could be contained 
within a single cluster with an effective radius of only 20 pc. However, reducing 
the estimated mass to $10^8~M_\odot$ yields an even narrower range of ages,  
$7 \times 10^5 \lta \tau \lta 5 \times 10^6$ yr. Clearly these {\UDHII}s are 
among the youngest extragalactic stellar clusters yet found.

It is interesting to compare the results of this analysis with those
determined above from the spectral models. For a cluster age of about
$3.6 \times 10^6$ yr and a mass of $2.5 \times 10^6 M_{\odot}$, we would
predict a radius for the Stromgren sphere surrounding the cluster of about 
2.9 pc, which is in excellent agreement with the radius $R_{\rm in}$ determined from 
fitting the SED (\S 3.2 and Table 3). We can, in fact, use the size of the ionized 
region to place additional constraints on the mass and age of the cluster. 
As expected from the above comparison of the radii, the resulting curve lies 
close to that obtained from the total ionization rate derived from the radio 
fluxes. Alternatively, if the mass and age of the cluster are known 
the comparison between $R_{\rm in}$ and the predicted Stromgren radius allows us to 
place limits on the amount dust contained within the ionized region.
For the age-mass pair used above, the fraction of ionizing photons
absorbed by dust within the Stromgren sphere, $f_d$, must be close to zero. However,
for younger and/or more massive clusters, $f_d$ rises rapidly: for an age
of $3.0 \times 10^6$ yr and a mass of $3.0 \times 10^6 M_{\odot}$, for
example, $f_d = 0.62$.

The lower mass limits found for the stellar clusters embedded in these {\UDHII}s 
are far larger than the typical
masses found for SSCs in He 2-10 (Johnson et al.\ 2000), NGC 1714 (Johnson et al.\
1999), and many other starburst galaxies (see Schweizer 1999; Whitmore 2000
and references within). The mass of the cluster in knot 4 is about a factor of
10 larger than that estimated for the most luminous SSC in He 2-10 (Johnson et al.\
2000); for the other knots, the estimated masses are factors of several larger than
that of the brightest optical SSC. In deriving our constraints on the stellar
cluster ages and masses, we have assumed that each {\UDHII} contains a single 
embedded cluster, with an initial mass function extending down to $0.1 M_{\odot}$. 
If the mass function extends down only to $1 M_{\odot}$ (the typical 
value assumed for mass estimates of SSCs), the stellar cluster masses would be 
reduced by a factor of 2.55. Furthermore, the extended morphology of knot 4 could 
be due to the superposition of several unresolved sources. Large {\UCHII} complexes 
in the Galaxy, such as W49A for example, consist of a conglomeration of several 
embedded clusters (De Pree et al.\ 1997; Smith et al.\ 2000). 
Whether a given {\UDHII} consists of a single embedded stellar cluster, or a
complex of several clusters, each surrounded by their own dust cocoons and residing
within a lower density intercluster medium, is unknown at present.
Because our mass and age limits are derived from the total observed
broadband fluxes of each {\UDHII}, they represent the luminosity-weighted (or 
mass-weighted) characteristics for the sum of the embedded clusters, if several are 
contained within an {\UDHII}. Hence individual subclusters would have smaller 
masses and could have a range of ages compared to our estimates. 
Nevertheless, the mass estimates we have derived under the assumption
of a single stellar cluster are within the observed range (albeit at the high end)
of the masses found for SSCs (e.g., Mengel et al.\ 2001) and nuclear stellar
clusters (B\"oker, van der Marel, \& Vacca 1999). 
Recent work by Johnson et al.\ (2001) suggests there is a continuum of sizes 
and masses of embedded extragalactic clusters, from those with only a few O7V* 
stars \footnote{O7V* is the number of ``equivalent'' O7V-type stars required
to create a given Lyman continuum flux; a single O7V* produces
$10^{49}$ Lyman continuum photons s$^{-1}$ (Vacca 1994).}, to
moderate sized clusters of a few hundred O7V* (which are 
observationally nearly identical to {\UCHII} complexes in the Galaxy
such as W49A), up to these giant {\UDHII}s, which may be proto-globular
clusters. 
The question of whether subclusters are present within {\UDHII}s could  
be settled by higher spatial resolution radio and infrared observations or possibly 
by interferometric observations in the $N$-band,
for example with the MIDI instrument on the VLTI (Leinert et al.\ 2000; 
Lopez et al.\ 2000).

\section{DISCUSSION}
\subsection{Comparison with Ultracompact H{\small II} Regions}

In Figure ~\ref{SEDs} we plot the estimated SEDs
for knots 1+2, 3, 4, and 5 seen in the $N$-band and radio images. 
Comparison with the SEDs of Galactic {\UCHII} regions (e.g., Wood \& Churchwell 1989)
reveals a striking similarity and reinforces the  
suggestion of KJ99 that the knots in He 2-10 are in fact the extragalactic
analogues of {\UCHII}s, albeit vastly scaled up in size.
Indeed, the inner dust temperature of 189 K found from our analysis of the SEDs of
the knots in He 2-10 (Table 3) is remarkably close to 
that predicted by the simple relation between radius and temperature
given by Wilner, Welch, \& Forster (1995) for Galactic {\UCHII}s, 
\begin{equation}
T_{dust}{\rm (K)} = 54\Bigl(\frac{L}{10^6 L_\odot}\Bigr)^{0.25}\Bigl(\frac{r}{1 {\rm pc}}\Bigr)^{-0.4}
\end{equation}
even though the lumosity is well beyond the range for which the relation was 
derived (see e.g., Wolfire \& Cassinelli 1986). 

Although the association of the radio knots in He 2-10 with Galactic {\UCHII} regions appears
appropriate, the two types of objects are not identical. In Figure ~\ref{ratio} we plot the
ratio of the observed fluxes for knot 4 to those of W49A after scaling the
latter to a distance of 9 Mpc. W49A is one of the most well-studied Galactic {\UCHII} 
complexes and has been resolved into at least 30 individual {\UCHII}s 
(De Pree et al.\ 1997; Conti
\& Blum 2001). We have taken the fluxes for W49A at the various frequencies
from the results given by Mezger, Schraml, \& Terzian (1967) and 
Ward-Thompson \& Robson (1990). Figure \ref{ratio} shows that flux ratio increases
steadily with frequency, from a value of about 11 at 6 cm, to well over 200
at 12 \mm. This suggests that the {\UDHII}s contain far more dust than {\UCHII}s
relative to their ionizing luminosity. 
The trend in the flux ratio could be due to different mean ages for the stellar clusters
within the two types of objects, as the ionizing luminosity from a cluster drops more 
rapidly than the bolometric luminosity as a cluster ages. However, such an explanation
would require the mean age of the clusters in W49A to be substantially younger than those
in the {\UDHII}s in He 2-10.
With ages $\lta {\rm few} \times 10^6$ yr, the clusters embedded with the {\UDHII}s 
are clearly very young. If the mean age of the clusters within W49A were larger than that 
for the {\UDHII}s, one would expect the flux ratio to decrease with frequency, opposite 
of the trend seen in Figure \ref{ratio}. 

Because the flux ratio in Figure \ref{ratio} continues to increase
from 100 to 12 \mm, the relative fraction of hot dust ($T_{dust} > 100$K) must be greater 
in the {\UDHII}s than in W49A. The increased fraction of hot dust in the {\UDHII}s in 
He 2-10 could be due to external heating from other starburst regions in the vicinty. 
Given the tremendous star formation activity occuring in the center of He 2-10, as well 
as the location of the knots close to, or on the edges of sites of vigorous star formation 
activity, this does not seem implausible. If external heating of the dust is significant,
the bolometric luminosity (and hence the lower mass limits) for the embedded stellar clusters 
would be substantially smaller than we have estimated. 
However, the superb match between 
the radio and the $N$-band morphologies, the rather spherical shapes of the $N$-band knots, and
the lack of any significant edge-brightening on the sides nearest the optical starburst
regions suggests that the SSCs detected in the UV/optical HST images do not
provide a substantial fraction of this heating. We can attempt to estimate their heating
contribution in the following way. If we assume that the optical SSCs and the {\UDHII}s
are at the same relative distance, we can use the effective radii of the {\UDHII}s 
measured from the $N$-band image to compute the solid angle subtended by any knot as
seen from the SSCs. Knot 4 has $R_{\rm eff} = 21$ pc and is located about 40 pc from
three SSCs seen in the UV and optical HST images of He 2-10. Assuming that these
SSCs have masses of $10^5~M_{\odot}$, we estimate the maximum available bolometric luminosity 
impinging on knot 4 from these sources is $1.6 \times 10^7 L_\odot$. This is less
than $\sim 1$\% of the total infrared luminosity of the knot.

Chini, Kr\"ugel, \& Wargau (1987) found a correlation between the bolometric
luminosity of Galactic {\UCHII}s and their total gas mass,

\begin{equation}
L_{bol}/L_\odot = 56 (M_{gas}/M_\odot)^{0.93}~~~~.
\end{equation}
Using the estimate of $M_{gas}$ derived from our fit to the SED of knot 4, we
find that the bolometric luminosity of this knot is factor of 3 larger than
predicted by the relation above. As discussed above, it is certainly possible 
that the optically thin models underestimate the total gas mass. Nevertheless, 
again we find that the {\UDHII}s are overluminous in the far infrared relative to 
{\UCHII}s. The {\UDHII}s also do not lie on the size-density relation found by 
Garay et al.\ (1993) and Kim \& Koo (2001) for Galactic {\UCHII}s 
[$n_e ({\rm cm}^{-3}) = 790(D/{\rm pc})^{-0.99}$]; 
they are either far too dense for their size (as measured by either their $R_{\rm eff}$ or
$R_{\rm in}$) or too large for their estimated electron density.

The results of these comparisons {\em may} indicate that the total dust mass required 
to form a stellar cluster of a given mass increases non-linearly with the cluster mass. 
This in turn would imply that the star formation efficiency varies 
with the mass of the embedded cluster and is not nearly as high in the {\UDHII}s as in the 
far less massive Galactic {\UCHII}s, which contain only one to a few massive O stars. 
However, given the uncertainties, and the numerous other possible explanations (both 
observational and physical) for the trend seen in Figure \ref{ratio},  
this suggestion is clearly only speculative.  


\subsection{Implications for the Infrared-Radio Correlation
and Ultraluminous Infrared Galaxies \label{ULIRGs}}

There is a well-known correlation between the infrared and radio
fluxes of galaxies over a wide range of Hubble types, first 
reported by van der Kruit (1971). This relation is typically
represented by $q$, the logarithm of the far-infrared(FIR)-to-radio flux ratio, given by
\begin{equation}
q~ \equiv~ {\rm log}~[(F_{\rm FIR}~/~3.75 \times 10^{12}{\rm Hz})~
/~S_{1.49 {\rm GHz}}],
\end{equation}
where $F_{\rm FIR}$ is an approximation to the total flux between $\sim 40$ and
$\sim 120$~$\mu$m given by
\begin{equation}
{\rm F_{\rm FIR}~(W~m^{-2})}~ \equiv~ 1.26 \times 10^{-14}~
(2.58 S_{60\mu m} + S_{100\mu m}),
\label{eq:FIR}
\end{equation}
where $S_{1.49 {\rm GHz}}$, $S_{60\mu m}$, and $S_{100\mu m}$ are
the flux densities at 1.49~GHz (in units of W~m$^2$~Hz$^{-1}$) and 
60$\mu$m and 100$\mu$m (in Jy), respectively.

Over a range of normal, irregular, and starburst galaxies
$q$ is very robust with a small standard deviation: $\langle q \rangle = 2.35 \pm 0.2$
for a sample of galaxies with $\log L_{\rm IR}/L_\odot > 10$
(see Sanders \& Mirabel 1996 and references therein). 
For starburst galaxies selected from the IRAS Bright Galaxy Survey with 
$\log L_{\rm FIR}/L_\odot > 11.25$,
Condon et al.\ (1991b) find an average value of $q = 2.34 \pm 0.19$. However, they
also find that the more infrared luminous galaxies tend to have larger 
values of $q$. We have compiled $q$ values for a sample of 19 starburst
galaxies with $\log L_{\rm FIR}/L_\odot > 11.7$ (corresponding to 
$\log L_{\rm IR}/L_\odot > 12$;
see Sanders \& Mirabel 1996 for a definition of $L_{\rm IR}$) taken from Condon et al.\
(1991) and Carilli \& Yun (2000). The mean and median $q$ values for this set of
Ultra-Luminous Infrared Galaxies (ULIRGs) are $2.55 \pm 0.34$ and 2.44, respectively.
Furthermore, those galaxies with high $q$ values tend to have compact radio 
sources.  Condon et al.\ (1991b) hypothesized that most 
ULIRGs are powered by dense, compact starbursts which 
are optically thick at 1.49~GHz, and whose radio flux density is therefore 
suppressed relative to the far-infrared flux. Correcting for free-free absorption at 1.49 GHz 
yields an even tighter FIR-radio correlation. Such dense, optically thick starburst
regions are exactly what our $N$-band (and radio) observations of He 2-10  
have revealed. The dusty infrared engine of this starburst
galaxy turns out to be the young, compact {\UDHII}s first identified
by KJ99, objects which are clearly related to nuclear stellar clusters, SSCs 
and, by extension, globular clusters.

It is interesting to estimate the $q$ for the extreme case of a galaxy composed 
entirely of {\UDHII}s. We estimated the fluxes at 1.49 GHz for 
each of the {\UDHII}s in He 2-10 from our model fits and derived $F_{\rm FIR}$ from the 
SED points using Eq. \ref{eq:FIR}. The resulting $q$ values are given in Table 2 and range 
from 3.7 (for knot 3) to 4.8 (for knot 4), far larger (as expected) than the typical 
values found 
for starbursts, and larger even than the typical values found for ULIRGs.
For comparison, we also calculated $q$ for W49A. 
From the values given by Ward-Thompson \& Robson (1990) for
$S_{60\mu m}$ and $S_{100\mu m}$, ${\rm FIR}=2.6 \times
10^{-9}$~W~m$^{-2}$ for W49A.  From Mezger, Schraml, \& Terzian (1967), we have
$F_\nu$(1.41~GHz)$=30 \pm 3$~Jy. These values yield $q = 3.36$ for W49A, which
is also significantly higher than the average value found in normal
galaxies, starbursts, and most ULIRGs.
(The fact that $q$ for W49A is smaller
than the values found for the {\UDHII}s in He 2-10 reflects the result
discussed above, namely that the {\UDHII}s are overluminous in the 
infrared relative to their radio flux, as compared with Galactic {\UCHII}s.)
 
Given the high $q$ value derived for the knots in He 2-10 and W49A, which 
are entirely dominated by extremely young and embedded massive stars, we
might also expect to find high $q$ values for the {\em galaxies} in which
we have detected {\UDHII}s.  Using the IRAS Point Source Catalog
fluxes, we find $F_{\rm FIR} = 1.12 \times
10^{-12}$~W~m$^{-2}$ for He 2-10 and $F_{\rm FIR} = 1.39 \times
10^{-12}$~W~m$^{-2}$ for NGC 5253.  From KJ99, $F_\nu$(1.42~GHz)$=70 \pm
7$~mJy for He~2-10, and from Condon et al.\ (1990; see also Turner, Ho \& 
Beck 1998), $F_\nu$(1.49~GHz)$=84
\pm 1$~mJy for NGC~5253. The resulting $q$ values are 2.63 and 2.65
for He~2-10 and NGC~5253, respectively, both significantly
higher than the average value derived for starburst galaxies, but
comparable with values found for ULIRGs. Even if we correct the radio
flux from He 2-10 for free-free self-absorption according to the prescription
given by Condon et al.\ (1991b), we find $q = 2.5$, well above the 
average value. These values demonstrate again that the total 
far-infrared spectra of these galaxies are dominated by the 
emission from the {\UDHII}s; in addition, the total radio fluxes must also have a 
component from the {\UDHII}s, whose spectra are powered 
by thermal free-free, rather than synchrotron, emission. They also suggest that $q$
values significantly larger than the mean of 2.34 may serve to
identify starburst galaxies which host energetically dominant 
{\UDHII}s. Furthermore, because high $q$ values are typical of
ULIRGs, it is possible that {\UDHII}s might have a significant role in powering the
infrared luminosities of these galaxies.

The fact that the overall $q$ value for He 2-10  does not
match those of the {\UDHII}s indicates that there is a significant contribution
from non-thermal synchrotron emission to the total radio flux. This is
confirmed by the radio spectral slope of $\alpha \approx -0.6$, measured by 
Allen et al.\ (1976), Mendez et al.\ (1999), and KJ99.
We can attempt to remove the non-thermal component and measure
$q_{therm}$, the $q$ value associated with only the thermal component;
this should reflect the $q$ values for the {\UDHII}s. Assuming that 
$S_\nu \sim \nu^{\alpha}$ and that the intrinsic power spectrum index
for synchroton emission is $\alpha = -0.8$, we estimated the thermal fraction of the
flux at 1.49 GHz from the observed radio spectral index and computed 
$q_{therm}$. We find $q_{therm} = 3.32$ for He 2-10.
Carrying out the same procedure for the sample of 16 ULIRGs discussed above for which 
radio spectral indices are available, we find a mean value of 
$q_{therm} = 3.08 \pm 0.31$ and a median value of 3.13. Given the approximate
nature of the separation into thermal and non-thermal components, these values are
reasonably close to those determined for the individual {\UDHII}s in He 2-10.

How common might {\UDHII}s be among starburst galaxies? Beck et al.\
(2000) have recently carried out radio mapping of a small sample of
nine Wolf-Rayet galaxies. In most of these, they find regions where the
spectral slope increases with frequency, a signature of optically thick
free-free emission. The estimated parameters for these regions are very
similar to those of the {\UDHII}s seen in He 2-10 and NGC 5253. Hence
{\UDHII}s appear to be quite common among this class of starbursts. Of
course, Wolf-Rayet galaxies are identified by the signatures of young 
massive stars in their integrated spectra (e.g., Vacca \& Conti 1992),
so perhaps it is not surprising to find {\UDHII}s in these objects.

\subsection{{\UDHII}s in Ultraluminous Infrared Galaxies?}

The above discussion naturally leads to the question of whether the 
FIR flux in LIRGs and ULIRGs can possibly be produced by {\UDHII}s. 
With $L_{\rm IR} = 5.4 \times 10^9 L_\odot$, He 2-10 clearly falls far short 
of the standard definition of a LIRG ($10^{11} < L_{\rm IR} < 10^{12} L_\odot$) 
or a ULIRG ($L_{\rm IR} > 10^{12} L_\odot$; e.g., Sanders \& Mirabel 1996). 
However, He~2-10 is also a dwarf galaxy with $M_{\rm H_2} \approx 5.6 \times 10^{8} M_\odot$ 
(Kobulnicky et al.\ 1995). Scaling its $L_{\rm IR}$ by the ratio of the 
molecular mass of a typical ULIRG, such as Arp~220 [$M_{\rm H_2} \approx 10^{10} M_\odot$; 
Radford, Solomon, \& Downes 1991; Scoville, Yun, \& Bryant 1997] to that of He 2-10, 
we would predict a `` massive He~2-10'' to have $L_{\rm IR} \approx 10^{11}$ and contain
approximately 100 {\UDHII}s. This would imply that $\gta 1000$ {\UDHII}s are 
needed to power the FIR emission from a ULIRG. Such a large number of {\UDHII}s, 
all with ages $\lta 10^6$ yr, and all located within a small radius of only a few 
hundred pc, perhaps seems unlikely. Nevertheless, the {\UDHII} hypothesis is attractive 
as it naturally explains the large $q$ values found for ULIRGs, as well as the dust 
temperatures and small sizes estimated for the starburst regions in these objects. 
Perhaps a single very massive {\UDHII} might be powering the FIR flux in these sources.
Furthermore, because the dust masses in ULIRGs are so much higher than that
found in He 2-10, it may well be possible that pressure confinement extends the 
lifetime of the {\UDHII} phase in these objects. 
Finally, the above comparison reveals that either ULIRGs are overluminous 
relative to their mass compared to objects like He 2-10, and hence their star
formation efficiencies are much higher, or that $M_{H_2}$ alone is
not a good tracer of the total material available for cluster formation. 

\subsection{Implications for the $F_{\rm FIR}/F_{1600}$ Correlation for Starburst
Galaxies}

Meurer et al.\ (1997, 1999) have shown that the ratio of the FIR flux to 
that at 1600 \AA\ is well correlated with the UV spectral slope, $\beta$. This
correlation has been interpreted as arising naturally from the fact that
the hot stars which produce the UV flux heat the dust surrounding the 
star formation site where they were born. The dust absorbs and reprocesses 
the UV flux and emits it in the far infrared. Meurer et al.\ (1995) find that
a foreground screen model and a fairly gray extinction curve provide the best 
fit to the observed correlation. 

It is interesting to determine where He 2-10 and NGC 5253 lie on the
$\beta$ vs $F_{\rm FIR}/F_{1600}$ diagram. Assuming that all of the observed flux in
the IRAS bands for He 2-10 is the region seen in our $N$-band image, and using the 
UV flux value for this region,
corrected for Galactic extinction, given by Johnson et al.\ (2000), 
we find that $\log (F_{\rm FIR}/F_{1600}) = 1.35$. The value of $\beta$, after
correcting the observed UV spectrum for Galactic extinction, is $\beta = -1.19$. 
These values place
He 2-10 slightly above the best fit line given by Meurer et al.\ (1999)
but not far from it, and certainly within the spread of data points used to
define the correlation. For NGC 5253, $\log (F_{\rm FIR}/F_{1600}) = 0.61$,
and $\beta = -1.43$ (Meurer et al.\ 1995), which place NGC 5253
very close to the best fit line given by Meurer et al.\ (1999). 

Since most, and perhaps almost all, of the FIR flux in He 2-10 and NGC 5253 is 
generated by the compact radio knots, and does {\em not} represent reprocessed
UV flux from the starbursts seen in the UV, optical, and near-infrared images 
of these galaxies,
the obvious question then is, Why do these objects appear to follow the
correlation between $F_{\rm FIR}/F_{1600}$ and $\beta$? The UV flux and the 
FIR emission are {\em not} generated by the same starburst regions. In fact,  
the {\UDHII}s individually cannot be plotted on this diagram at all, because 
they have no detectable UV flux. 

Given that {\UDHII}s may well be quite common in starburst galaxies, and hence
may dominate the FIR flux of these objects, it is surprising that the
correlation demonstrated by Meurer et al.\ (1999) exists at all. In fact, 
it could be argued that this correlation indicates that {\UDHII}s cannot
be common in starbursts or that their emission cannot dominate the FIR fluxes. 
However, the results presented by Beck et al.\ (2000), 
discussed above, directly contradict this argument. Alternatively,
since at least some of the observed FIR flux {\em must}
be due to dust heated by the UV-visible starbursts, although most of the FIR
flux arises from {\UDHII}s, one might expect that the correlation shown by Meurer et al.\ 
(1997, 1999) should actually represent a {\em lower envelope} for an ensemble of starbursts.
The fact that most ULIRGs do appear to lie above the correlation reinforces this suggestion
(Meurer et al.\ 2000). If this is indeed the case, then perhaps
much of the scatter seen in the $\beta$ vs $F_{\rm FIR}/F_{1600}$ correlation
presented by Meurer et al.\ (1997, 1999) might be eliminated if the contribution
of {\UDHII}s could be removed from the total FIR flux. On the other hand, the distribution
of points around the correlation does not exhibit any substantial asymmetry toward
higher values of $F_{\rm FIR}/F_{1600}$, as one would expect if this relation is
actually a lower envelope for starbursts in general. 

In any case, if most of the FIR flux in starbursts is indeed due to {\UDHII}s, this 
immediately implies that the {\it observed} FIR and 1600 \AA\ fluxes are 
not causally connected in the straightforward manner suggested by the simple foreground
dust screen reddening/reprocessing picture presented by Meurer et al.\ (1995). 
Rather, the correlation between $F_{\rm FIR}/F_{1600}$ and $\beta$ might instead 
represent merely a trend in the dust content among starburst galaxies: starbursts 
containing more numerous {\UDHII}s are simply dustier in general. 
Hence, the reddening of the observed UV continuum appears to correlate with the 
$F_{\rm FIR}/F_{1600}$ ratio, 
even though the {\em observed} fluxes in the FIR and UV are not directly coupled.

The observed correlation in the $\beta$ vs $F_{\rm FIR}/F_{1600}$ diagram
is naturally produced in the model presented by Charlot \& Fall
(2000) for dust absorption in starbursts. Charlot \& Fall (2000) include
the finite lifetime and optical depth of stellar birth clouds in their
attempt to reproduce the inferred extinction law and the observed correlations 
of parameters in starburst galaxies. They find that, under the assumption of 
constant star formation, the presence and finite lifetimes ($\lta 10^7$ yr) 
of these birth clouds are key to reproducing 
the empirical
observation that the reddening derived from the $H\alpha/H\beta$ is generally 
larger than that derived from the UV continuum. In this model, the ionizing 
photons responsible for the H lines arise primarily from hot stars deeply embedded 
in optically thick birth clouds, the UV arises from older stellar clusters
whose birth clouds have dispersed, and the infrared emission is produced
by dust reprocessing of non-ionizing UV flux from both populations in both
environments (birth clouds and
ambient ISM). However, most of the infrared emission (typically $\gta 60$\%; S.\
Charlot, private communication) is generated by the dust reprocessing in the birth clouds. 
The observed $F_{\rm FIR}/F_{1600}$ vs. $\beta$ correlation arises
in this model as a result of a trend in the overall dust content of starbursts,
as suggested by our findings, rather than extinction and reprocessing from a foreground 
screen of dust. 
Our Gemini results, in which $60$\% of the FIR flux is produced by young stellar
clusters deeply embedded in their birth clouds, provide strong support for this model.

\section{Summary and Conclusions}

We have obtained $N$-band (10.8 $\mu$m) images of the starburst galaxy He 2-10 with
OSCIR and the Gemini North Telescope. These images reveal the presence of four
bright, compact knots (with effective radii on the order of 10-20 pc) which have 
no counterparts in the HST UV, HST optical, or
ground-based near-infrared ($J$, $H$, and $K'$) images. However, the $N$-band 
morphology is extremely 
well-correlated with the radio maps presented by KJ99. Therefore, we identify 
these knots with the {\UDHII} regions found by KJ99. The four knots comprise
about $60$ \% of the total $N$-band flux from He 2-10. Assuming the morphology
seen in the $N$-band image persists to longer wavelengths (e.g., 100 $\mu$m),
we have scaled the observed IRAS fluxes from the galaxy by the
fraction each knot contributes to the total $N$-band flux and combined the resulting 
values with the $N$-band and radio fluxes to generate SEDs for the individual knots.
The SEDs bear a remarkable resemblance to those of Galactic
{\UCHII} regions. We have constructed simple models to match the SEDs and 
have derived physical parameters for each knot. We find that the electron
densities in the ionized regions surrounding the stellar clusters are on the 
order of $10^3-10^4$ cm$^{-3}$, and radii of these regions are between
2 and 5 pc. The masses of the surrounding dust shells are about $10^6-10^7~M_\odot$. The 
radio and IR fluxes also allow us to place constraints on the properties of the 
embedded stellar clusters. For the brightest knot we find a mass 
$M_{Cluster} > 2.5 \times 10^6~M_\odot$ and an age $\tau < 3.6 \times 10^6$ yr.
The stellar clusters within all of the knots must have ages less than $\sim 5  \times 10^6$ yr
and masses greater than $\sim 5  \times 10^6 M_{\odot}$.
Hence, we confirm the suggestion of KJ99 that the radio
knots represent deeply embedded SSCs, and may be the youngest examples
yet observed.

While the SEDs of the {\UDHII}s appear to be similar to Galactic {\UCHII}s,
a comparison between the brightest $N$-band knot (\# 4) and the Galactic source W49A
reveals that knot 4 emits far more IR flux relative to its radio flux than W49A. This
suggests that, compared to Galactic {\UCHII}s, knot 4 contains far more dust relative 
to the ionizing luminosity of the embedded cluster. In addition,
the fraction of hot dust is larger in this knot than in W49A. Furthermore, the {\UDHII}s
do not follow the size-density relation found for Galactic {\UCHII}s. 
It is not clear what causes these differences.
It is also not certain whether knot 4 consists of a single embedded stellar
cluster or of several clusters, each surrounded by their own cocoons. Higher
resolution radio and infrared observations would be useful to answer this question. 

The logarithmic ratio of the radio to far-infrared flux
densities, $q$, for the {\UDHII}s in He 2-10 is $\sim 4$.
For both He 2-10 as a whole and NGC 5253, another nearby starburst
known to host {\UDHII}s, $q \approx 2.6$. 
The latter $q$ value is significantly larger than the
average $q=2.35$ found for normal galaxies, but comparable to the
values of $q$ found for ultraluminous infrared galaxies.
We suggest that large $q$ values for starburst galaxies may indicate that a significant
fraction of the far-infrared flux may arise from dust emission from {\UDHII}s.

Recent radio surveys seem to indicate that such {\UDHII}s may be common among 
starburst galaxies. If this is indeed the case, a significant
fraction of the star formation in these galaxies is occuring within regions that 
are completely hidden to UV, optical, and even near-infrared observations. This 
in turn raises important questions about the origin and meaning of the empirical 
$F_{\rm FIR}/F_{1600}$ correlation observed for samples of starbursts. Attempts to
address these issues would benefit from a much larger sample of $N$-band 
observations of nearby starburst galaxies.

%


\acknowledgments 

This paper is based on observations obtained with the
mid-infrared camera OSCIR, developed by the University of Florida with
support from the National Aeronautics and Space Administration, and
operated jointly by Gemini and the University of Florida Infrared
Astrophysics Group. The authors would like to thank Chip Kobulnicky
for kindly providing the 1.3 cm fluxes and discussions regarding {\UDHII}s. 
W.D.V. appreciates stimulating discussions with S.\ Charlot about dust and extinction
models and D.\ Sanders and D.\ Rigoupoulo about ULIRGs. K.E.J. thanks 
R\'emy Indebetouw for help with computer software and useful conversations.  
K.E.J. is pleased to acknowledge support for this work provided by NASA through 
a Graduate Student Researchers Fellowship. P.S.C. appreciates continuous support 
from the National Science Foundation.


\ifsubmode\else
\baselineskip=10pt
\fi


\clearpage


\ifsubmode\else
\baselineskip=14pt
\fi


\newcommand{\figcapNimage}{$N$-band Gemini image of He~2-10. North is up and East is 
left. The field of view is $9.43'' \times 9.43''$; each pixel is $0.089''$. \label{Nimage}}

\newcommand{\figcapNIRimage}{$K'$-band images of He~2-10. North is up and 
East is left. The field of view and position are the same as in Figure 1. 
\label{Kimage}}

\newcommand{\figcapNRadio}{$N$-band Gemini image of He~2-10 overlaid with the 2~cm 
radio contours from KJ99.  Four of the five radio {\UDHII}s are also strong 10.8$\mu$m 
sources. The westernmost $N$-band knot may be composed of two {\UDHII}s seen in
the radio map.\label{Nradio}}

\newcommand{\figcapNoptical}{$N$-band contours overlaid on the optical V-band image
from HST (Johnson et al.\ 2000).  The mid-IR morphology is not correlated
with the regions of star formation apparent in the ultraviolet and optical bands.
\label{Noptical}}

\newcommand{\figcapSEDs}{The estimated SEDs for the four {\UDHII}s detected
with Gemini. The $N$-band fluxes (solid circles) are from this work. The radio points
(open squares) are from Kobulnicky (private communication; 1.3 cm) and KJ99
(2 cm and 6~cm). The fluxes in the IRAS bands (12, 25, 60, and 100 $\mu$m; open 
circles) have been estimated as described in the text.  \label{SEDs}}

\newcommand{\figcapSEDmodel}{The SED of knot 4 (data points) with the model SED described
in the text (solid line). The dashed line represents the dust emission. \label{SEDmodel}}

\newcommand{\figcaptemplot}{The temperature profile found for knot 4. 
The adopted functional form is $\log T \sim (r/R_{\rm in})^{-\delta}$; the best
fit to the SED yields $\delta = 0.11$. \label{templot}}

\newcommand{\figcapmtplot}{Age versus Mass plot for the stellar cluster embedded in 
knot 4. The blue line represents the constraints placed by the $L_{bol}$ for knot 4; 
the region below this line is excluded. The red line outlines the region excluded by 
the lack of a SN signature in the radio SED. The green line represents the constraint
given by various estimates of the ionizing flux estimated from the radio
spectrum. The diagonal purple line is the age constraint corresponding to a size of
the ionized region of 2.9 pc for $n_0 = 1.8\times 10^6$ (Equation \ref{size}). If the
{\UDHII} is in pressure equilibrium, the region below this line is excluded. \label{mtplot}}

\newcommand{\figcapratio}{Ratio of the observed fluxes of knot 4 in He 2-10 to those
of W49A. \label{ratio}}

\clearpage


\ifsubmode
\figcaption{\figcapNimage}
\figcaption{\figcapNIRimage}
\figcaption{\figcapNoptical}
\figcaption{\figcapNRadio}
\figcaption{\figcapSEDs}
\figcaption{\figcapSEDmodel}
\figcaption{\figcaptemplot}
\figcaption{\figcapmtplot}
\figcaption{\figcapratio}
\clearpage
\else\printfigtrue\fi

\ifprintfig


\clearpage
\begin{figure}
\plotone{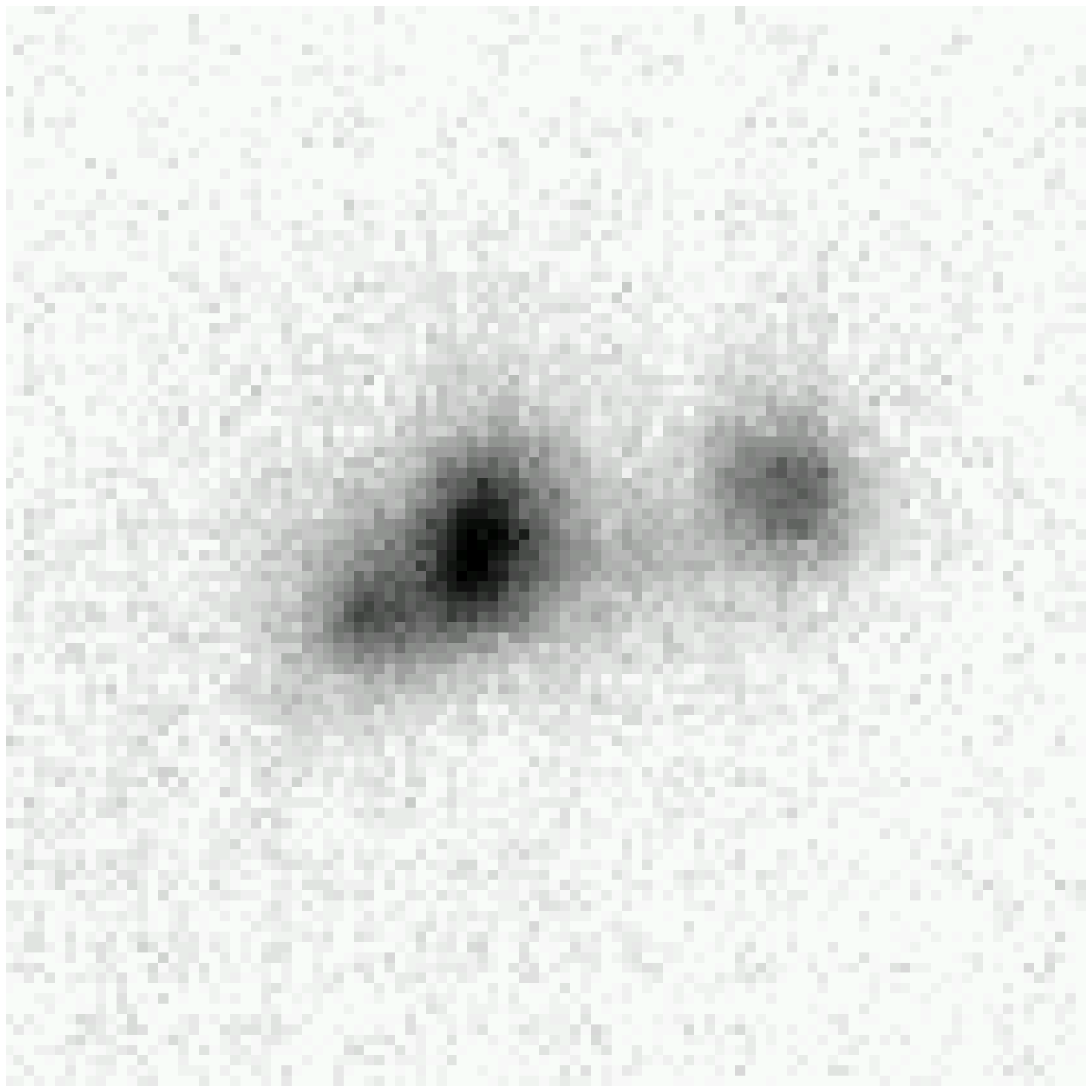}
\ifsubmode
\vskip3.0truecm
\addtocounter{figure}{1}
\centerline{Figure~\thefigure}
\else\vskip-0.3truecm\caption{\figcapNimage}\fi
\end{figure}


\clearpage
\begin{figure}
\plotone{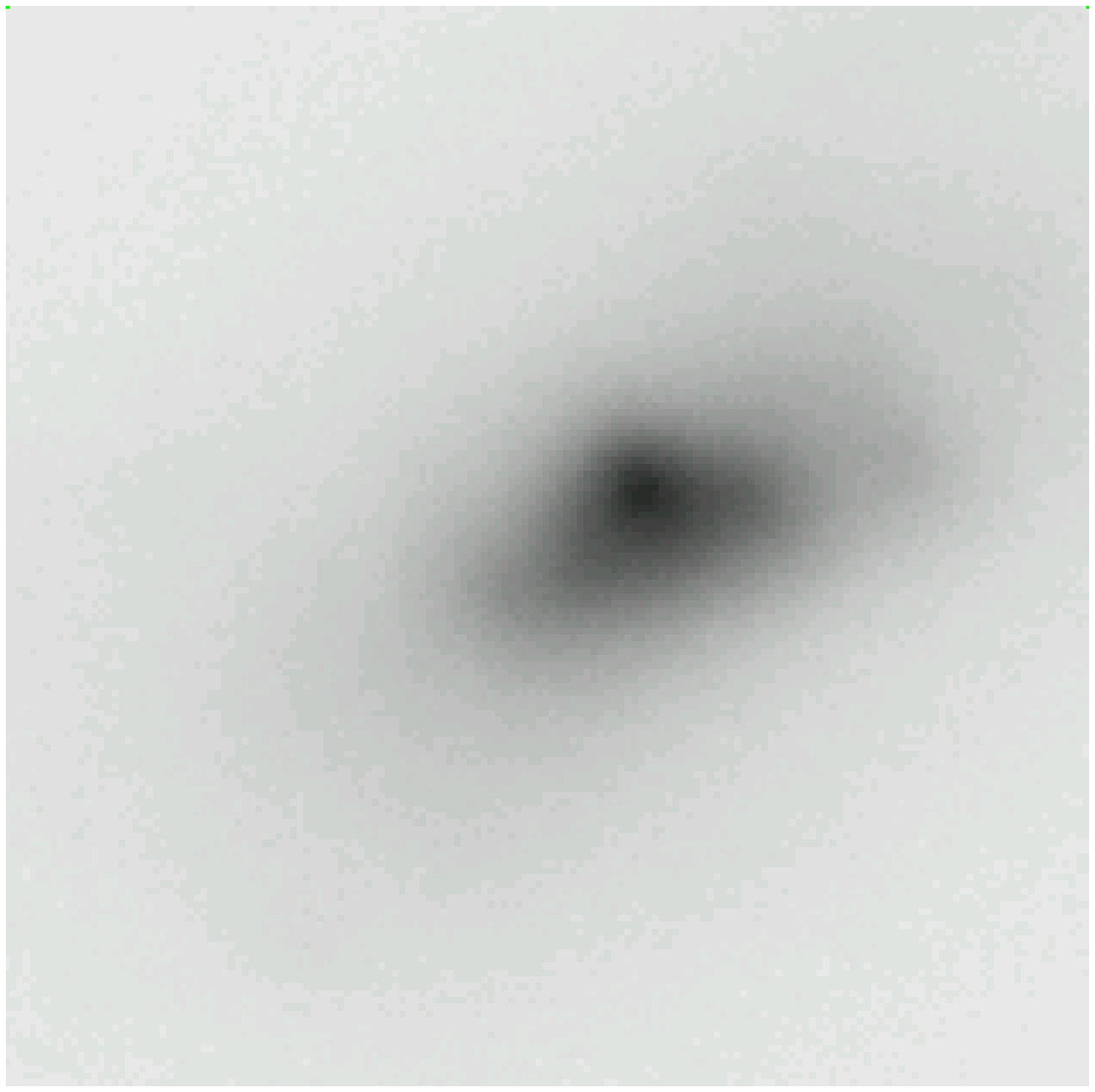}
\ifsubmode
\vskip3.0truecm
\setcounter{figure}{0}
\addtocounter{figure}{1}
\centerline{Figure~\thefigure}
\else\vskip-0.3truecm\caption{\figcapNIRimage}\fi
\end{figure}


\clearpage
\begin{figure}
\plotone{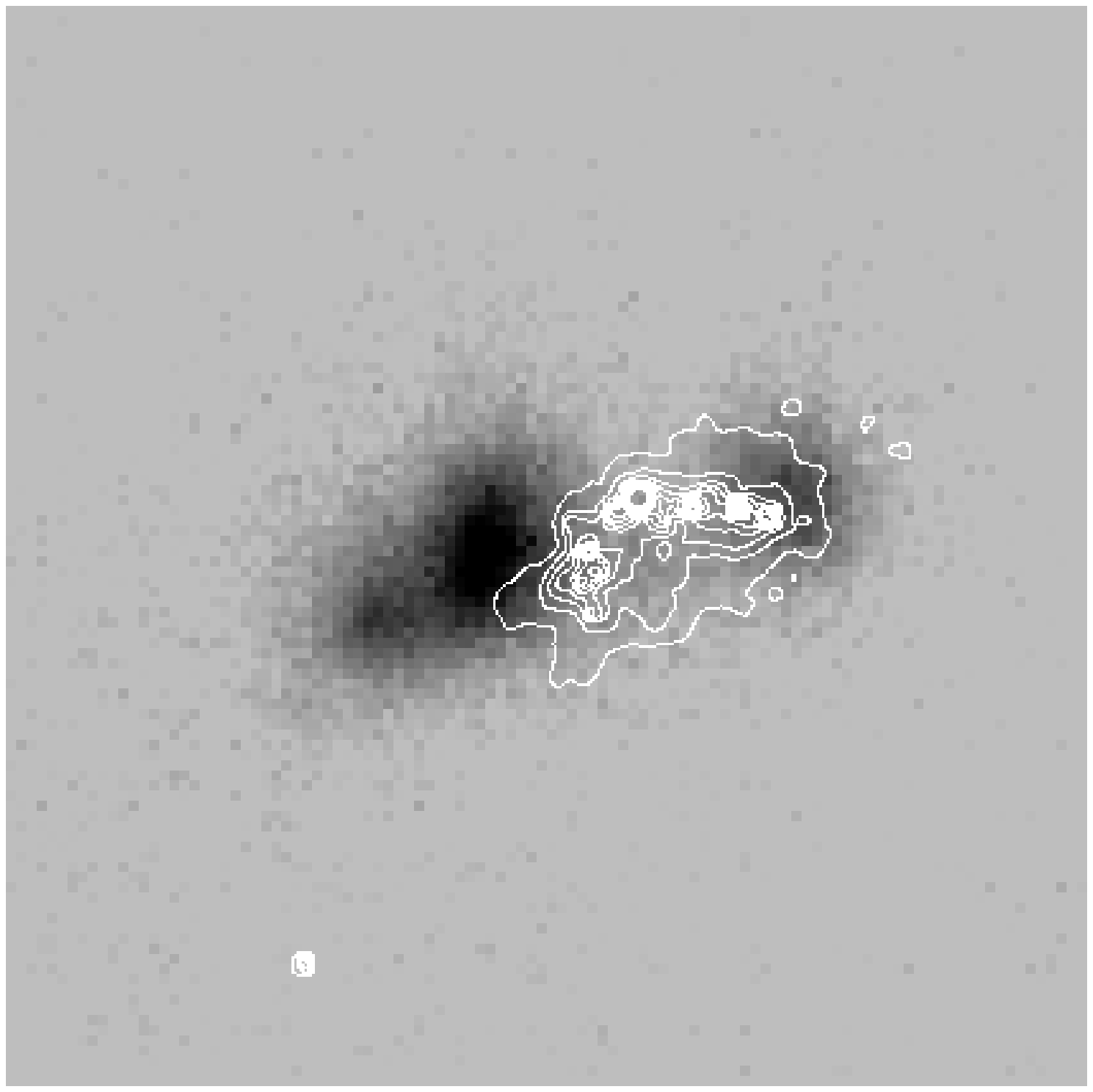}
\ifsubmode
\vskip3.0truecm
\addtocounter{figure}{1}
\centerline{Figure~\thefigure}
\else\vskip-0.3truecm\caption{\figcapNoptical}\fi
\end{figure}


\clearpage
\begin{figure}
\plotone{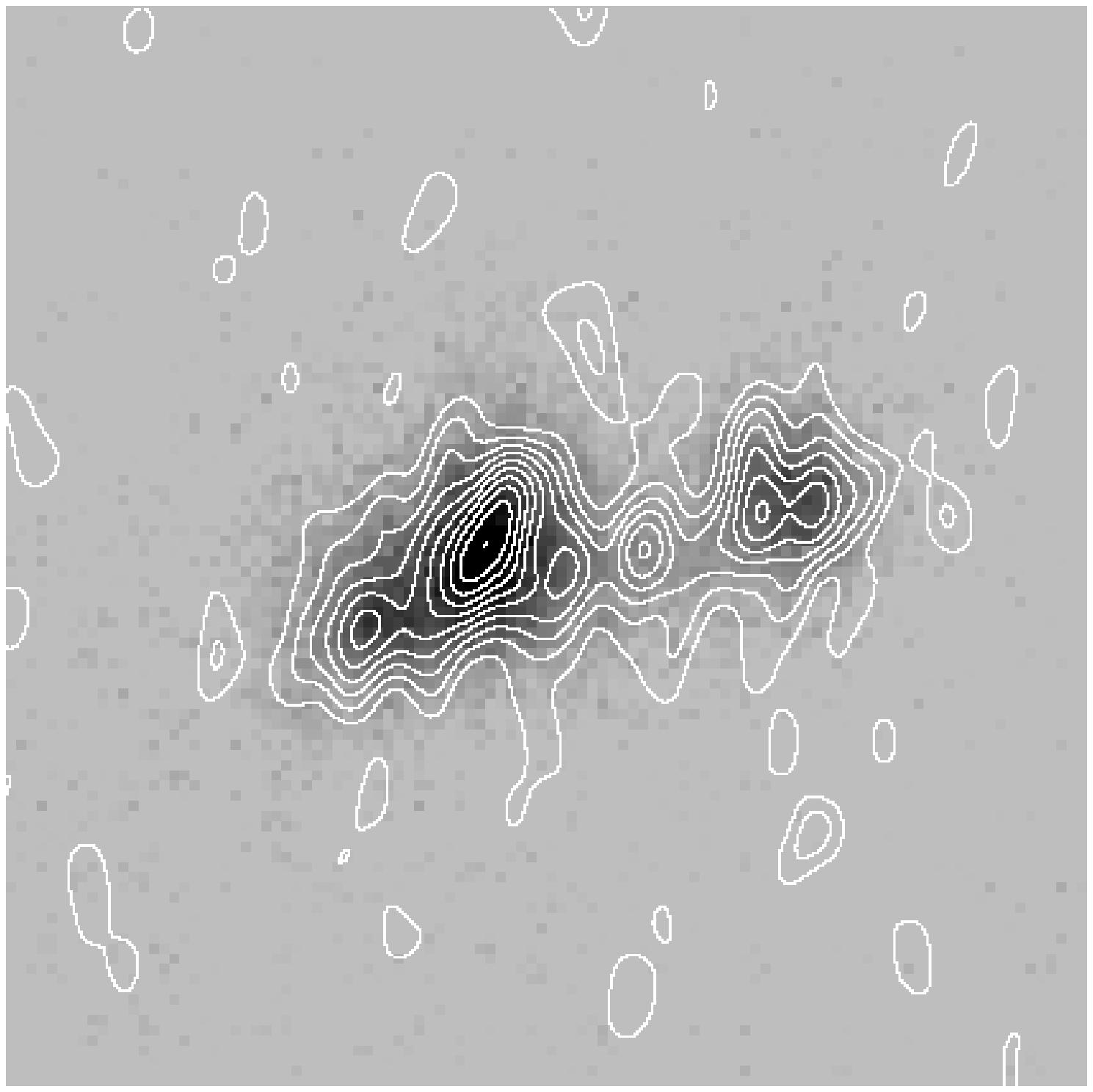}
\ifsubmode
\vskip3.0truecm
\addtocounter{figure}{1}
\centerline{Figure~\thefigure}
\else\vskip-0.3truecm\caption{\figcapNRadio}\fi
\end{figure}


\clearpage
\begin{figure}
\plotone{Vacca.fig5.ps}
\ifsubmode
\vskip3.0truecm
\addtocounter{figure}{1}
\centerline{Figure~\thefigure}
\else\vskip-0.3truecm\caption{\figcapSEDs}\fi
\end{figure}


\clearpage
\begin{figure}
\epsscale{0.8}
\plotone{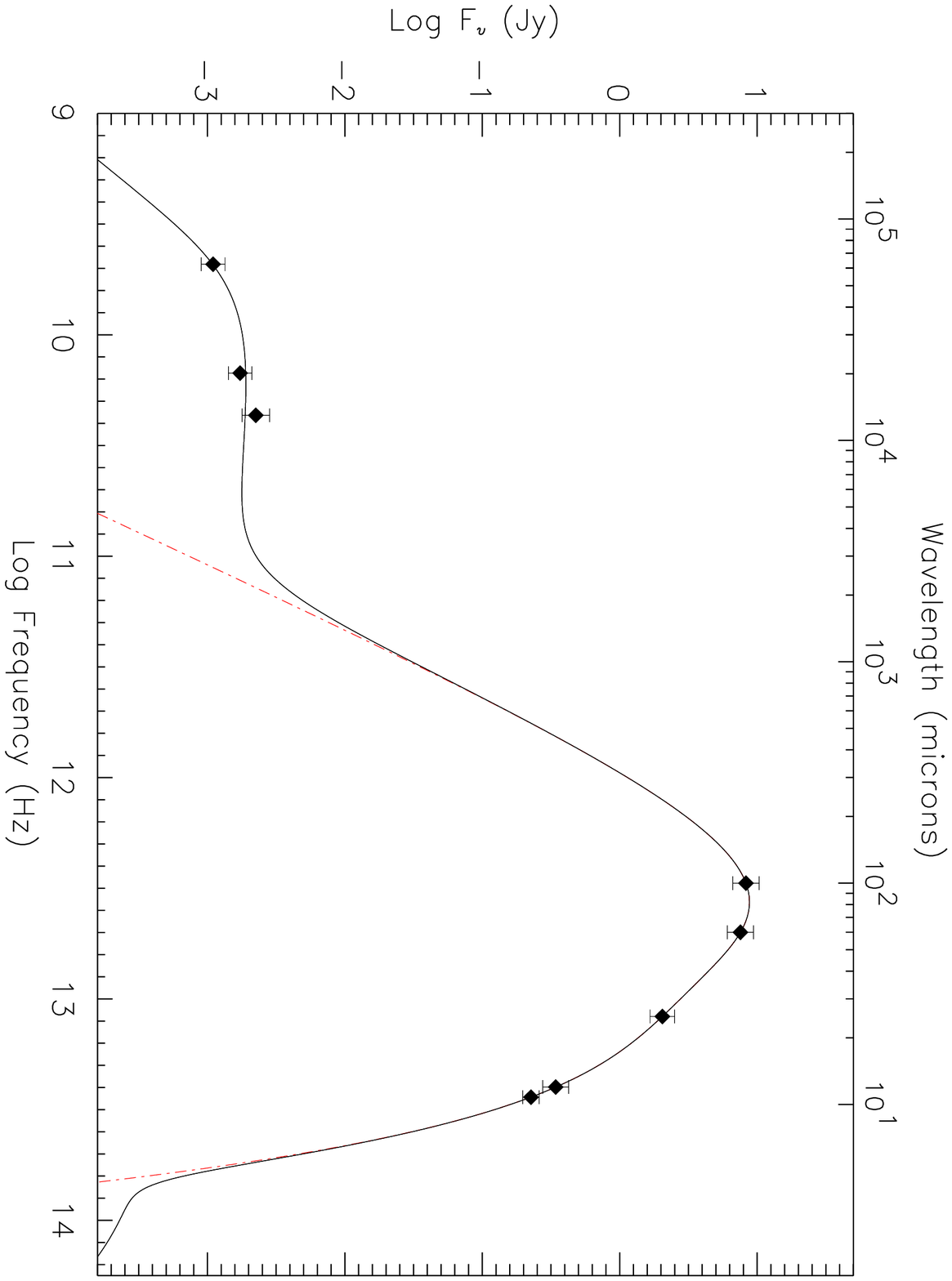}
\ifsubmode
\vskip3.0truecm
\addtocounter{figure}{1}
\centerline{Figure~\thefigure}
\else\vskip-0.3truecm\caption{\figcapSEDmodel}\fi
\end{figure}


\clearpage
\begin{figure}
\epsscale{0.8}
\plotone{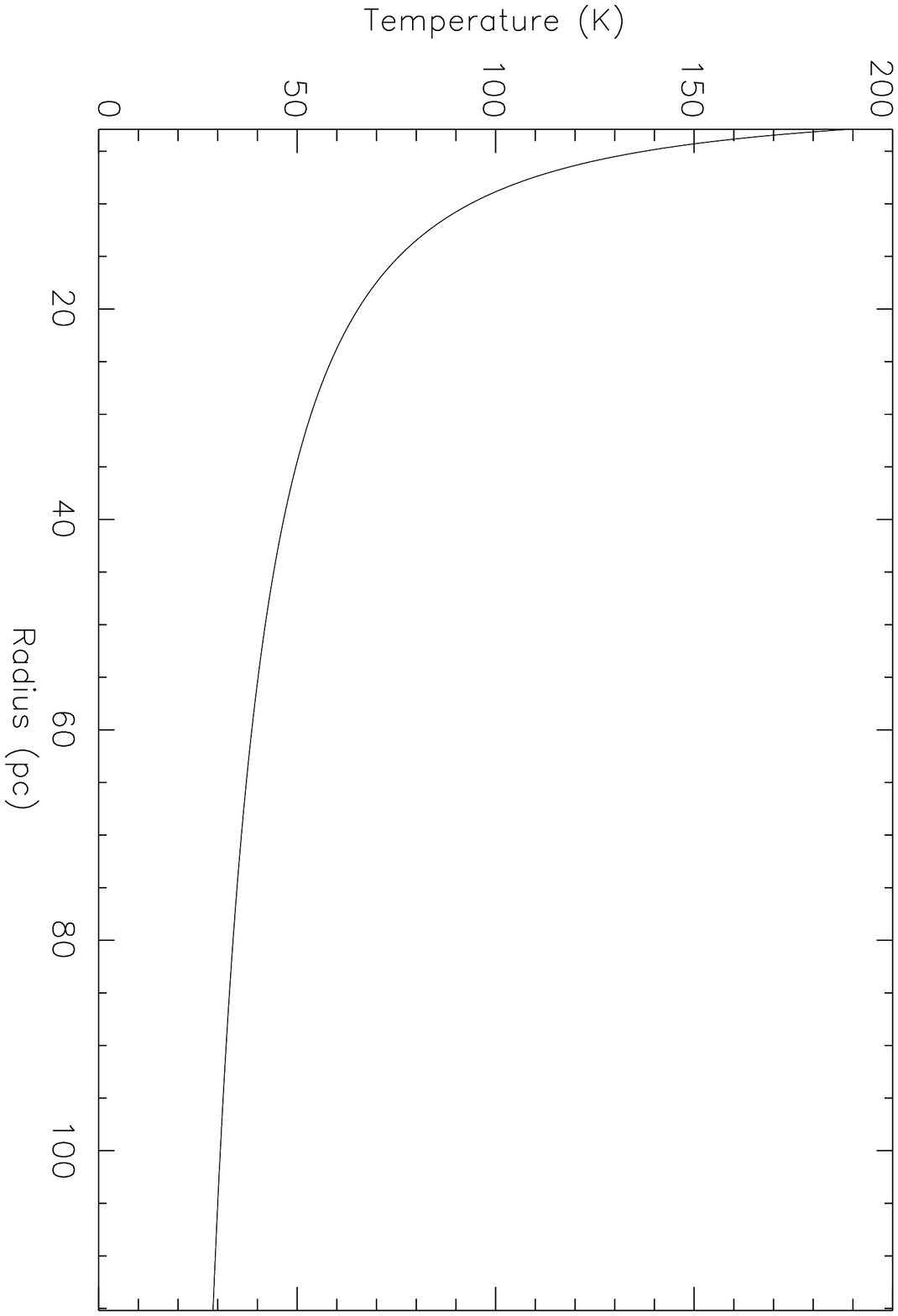}
\ifsubmode
\vskip3.0truecm
\addtocounter{figure}{1}
\centerline{Figure~\thefigure}
\else\vskip-0.3truecm\caption{\figcaptemplot}\fi
\end{figure}


\clearpage
\begin{figure}
\epsscale{0.8}
\plotone{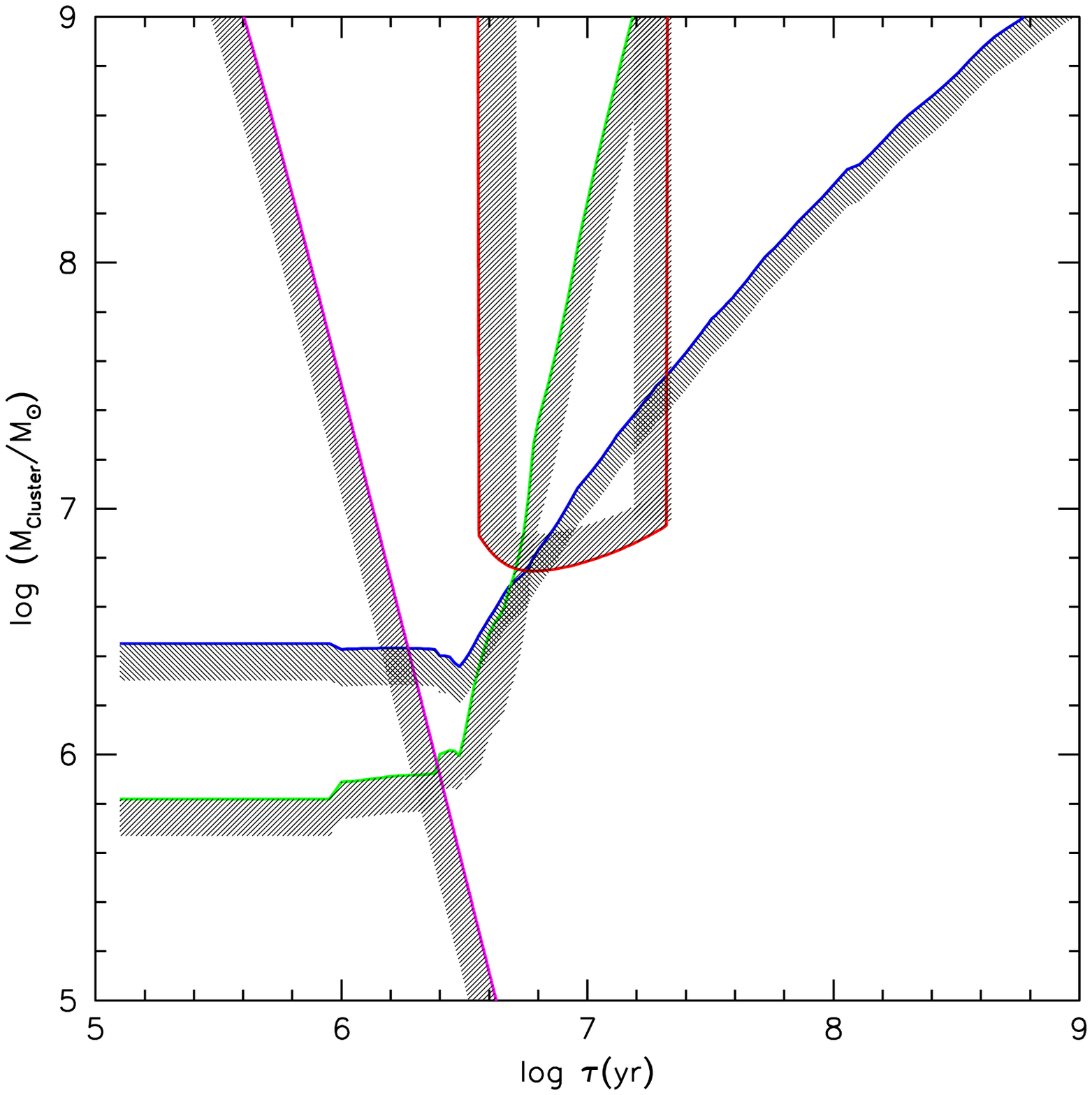}
\ifsubmode
\vskip3.0truecm
\addtocounter{figure}{1}
\centerline{Figure~\thefigure}
\else\vskip-0.3truecm\caption{\figcapmtplot}\fi
\end{figure}


\clearpage
\begin{figure}
\epsscale{0.8}
\plotone{Vacca.fig9.ps}
\ifsubmode
\vskip3.0truecm
\addtocounter{figure}{1}
\centerline{Figure~\thefigure}
\else\vskip-0.3truecm\caption{\figcapratio}\fi
\end{figure}


\fi 


\clearpage
\ifsubmode\pagestyle{empty}\fi


\begin{deluxetable}{llll}
\tablenum{1}
\tabletypesize{\small}
\tablecaption{$N$-band fluxes, magnitudes, and effective radii \label{mag.tab}}
\tablewidth{0pt}
\tablehead{
\colhead{Object} & 
\colhead{Flux (mJy)} &
\colhead{N mag} & 
\colhead{$R_{\rm eff}$~(pc)} \\ 
}
\startdata
1 \& 2& $101  \pm 17$ & $6.43 \pm 0.19$ & 15.7 \\
3     & ~~$41 \pm ~9$ & $7.42 \pm 0.25$ & 17.1 \\
4     & $226  \pm 31$ & $5.56 \pm 0.15$ & 21.0 \\
5     & ~$65  \pm 13$ & $6.92 \pm 0.21$ & 10.3 \\
\enddata
\end{deluxetable}


\begin{deluxetable}{lcccc}
\tablenum{2}
\tabletypesize{\small}
\tablecaption{Estimated fluxes for the radio/$N$-band knots \label{flux.tab}}
\tablewidth{0pt}
\tablehead{
\colhead{Wavelength} &
\multicolumn{4}{c}{~~~~~~~~~~~Flux (mJy)~~~~~~~~~~~~} \\
\multicolumn{5}{c}{~} \\
\colhead{~ } &
\colhead{knots 1\&2} &
\colhead{knot 3} &
\colhead{knot 4} &
\colhead{knot 5} \\
}
\startdata
$10.8~ \mu$m & ~101 $\pm$ ~~17 & ~~41 $\pm$ ~~9  & ~226 $\pm$ ~~31 & ~~65 $\pm$ ~13 \\
$12~ \mu$m   & ~153 $\pm$ ~~36 & ~~62 $\pm$ ~17  & ~342 $\pm$ ~~74 & ~~98 $\pm$ ~25 \\
$25~ \mu$m   & ~913 $\pm$ ~208 & ~371 $\pm$ ~99  & 2043 $\pm$ ~421 & ~588 $\pm$ 149 \\
$60~ \mu$m   & 3378 $\pm$ ~811 & 1371 $\pm$ 383  & 7558 $\pm$ 1663 & 2174 $\pm$ 574 \\
$100~ \mu$m  & 3703 $\pm$ ~889 & 1503 $\pm$ 419  & 8287 $\pm$ 1823 & 2383 $\pm$ 629 \\
$1.3$ cm     & 1.40 $\pm$ 0.21 & ...             & 2.25 $\pm$ 0.52 & 1.25 $\pm$ 0.29 \\ 
$2$ cm       & 2.00 $\pm$ 0.28 & 0.89 $\pm$ 0.18 & 1.73 $\pm$ 0.34 & 1.04 $\pm$ 0.21 \\
$6$ cm       & 1.38 $\pm$ 0.19 & 0.89 $\pm$ 0.18 & 1.10 $\pm$ 0.22 & 0.57 $\pm$ 0.12 \\
$\alpha^6_2~^a$ & $+0.33^b$    & $~0.00$         & $+0.38$         & $+0.51$         \\
$q^c$        & 4.2             & 3.7             & 4.8             & 4.6             \\
\enddata
\tablenotetext{\it a}{Spectral index between 2 and 6 cm ($S_{\nu} \propto \nu^{\alpha}$) from KJ99.}
\tablenotetext{\it b}{Mean spectral index for knots 1 and 2.}
\tablenotetext{\it c}{$q~=~\log~[(F_{\rm FIR}~/~3.75 \times 10^{12}{\rm Hz})/S_{1.49{\rm GHz}}]$}
\end{deluxetable}


\begin{deluxetable}{lcccc}
\tablenum{3}
\tabletypesize{\small}
\tablecaption{Estimated parameters for the {\UDHII} cocoons \label{pars}}
\tablewidth{0pt}
\tablehead{
\colhead{Parameter} &
\colhead{knots 1\&2} &
\colhead{knot 3} &
\colhead{knot 4} &
\colhead{knot 5} \\
}
\startdata
\multicolumn{5}{l}{Ionized Region} \\
$n_e$ (cm$^{-3}$)      & 2240              & 1400              & 4290              & 5720              \\
$R_{in}$ (pc)          & 4.2               & 4.7               & 2.9               & 2.0               \\
$M_{H^+}/M_\odot$      & $1.7 \times 10^4$ & $1.5 \times 10^4$ & $1.1 \times 10^4$ & $4.9 \times 10^3$ \\
EM (cm$^{-6}$ pc)$^a$  & $4.2 \times 10^7$ & $1.8 \times 10^7$ & $1.1 \times 10^8$ & $1.3 \times 10^8$ \\
\multicolumn{5}{c}{ ~ } \\
\multicolumn{5}{l}{Dust Shell} \\
$R_{in}$ (pc)          & 4.2               & 4.7               & 2.9               & 2.0               \\
$R_{out}$ (pc)         & 170               & 198               & 115               & 91                \\
$T_{in}$ (K)           & 190               & 192               & 189               & 194               \\ 
$T_{out}$ (K)          & 29                & 28                & 29                & 28                \\
$L_{\rm IR}/L_\odot$   & $7.9 \times 10^8$ & $3.2 \times 10^8$ & $1.8 \times 10^9$ & $5.1 \times 10^8$ \\
$M_{\rm shell}/M_\odot$& $5.4 \times 10^6$ & $2.2 \times 10^6$ & $1.2 \times 10^7$ & $3.6 \times 10^6$ \\
\enddata
\tablenotetext{\it a}{Emission measure, $EM = \int n_e dr$, through the center of the ionized region.}
\end{deluxetable}


\clearpage


\end{document}